\documentclass[12pt,draftclsnofoot,onecolumn]{IEEEtran}

\usepackage{amsmath,amssymb,dsfont,stfloats,color,url,bbm}
\usepackage[pdftex]{graphicx}
\usepackage{subfigure}
\usepackage{cite}
\usepackage[squaren]{SIunits} 
\usepackage[boxruled,vlined,linesnumbered,inoutnumbered]{algorithm2e}	
\IncMargin{0.3em}
\SetKwInput{KwIn}{\underline{Input}}
\SetKwInput{KwOut}{\underline{Output}}
\SetKwInput{defi}{\underline{Definitions}}
\SetKwInput{defalg}{\underline{Definitions/Algorithm}}
\SetKwInput{alg}{\underline{Algorithm}}
\DontPrintSemicolon
\newlength{\commentWidth}
\DeclareGraphicsExtensions{.eps,.pdf,.png,.jpg,.gif,.jpeg,.pstex}
\newtheorem{theorem}{Theorem}

\newtheorem{lemma}{Lemma}

\newtheorem{example}{Example}

\newtheorem{definition}{Definition}
\newtheorem{remark}{Remark}

\newcommand{\m}[1]{\mathrm{#1}}	
\newcommand{\ve}[1]{\mathbf{#1}}
\newcommand{\ma}[1]{\mathbf{#1}}
\newcommand{\map}[1]{\left(\mathbf{#1}\right)}
\newcommand{\mr}[1]{(\ref{#1})}	
\newcommand{\un}[2]{\unit{#1}{#2}}	
\newcommand{\re}{\operatorname{\Re }}
\newcommand{\im}{\operatorname{\Im }}

\newcommand{\R}{\operatorname{\mathbb{R}}}
\newcommand{\C}{\operatorname{\mathbb{C}}}

\newfont{\bbb}{msbm10 scaled 700}

\newfont{\bb}{msbm10 scaled 1100}
\newcommand{\CC}{\mbox{\bb C}}

\newcommand{\RR}{\mbox{\bb R}}

\newcommand{\ZZ}{\mbox{\bb Z}}

\newcommand{\EE}{\mbox{\bb E}}

\newcommand{\HH}{\mbox{\bb H}}

\newcommand{\av}{{\bf a}}
\newcommand{\bv}{{\bf b}}

\newcommand{\dv}{{\bf d}}
\newcommand{\ev}{{\bf e}}

\newcommand{\gv}{{\bf g}}
\newcommand{\hv}{{\bf h}}

\newcommand{\pv}{{\bf p}}
\newcommand{\qv}{{\bf q}}
\newcommand{\rv}{{\bf r}}
\newcommand{\sv}{{\bf s}}

\newcommand{\uv}{{\bf u}}
\newcommand{\wv}{{\bf w}}
\newcommand{\vv}{{\bf v}}
\newcommand{\xv}{{\bf x}}
\newcommand{\yv}{{\bf y}}
\newcommand{\zv}{{\bf z}}
\newcommand{\zerov}{{\bf 0}}
\newcommand{\onev}{{\bf 1}}

\newcommand{\Am}{{\bf A}}
\newcommand{\Bm}{{\bf B}}
\newcommand{\Cm}{{\bf C}}
\newcommand{\Dm}{{\bf D}}

\newcommand{\Fm}{{\bf F}}
\newcommand{\Gm}{{\bf G}}
\newcommand{\Hm}{{\bf H}}
\newcommand{\Id}{{\bf I}}

\newcommand{\Pm}{{\bf P}}

\newcommand{\Rm}{{\bf R}}

\newcommand{\Wm}{{\bf W}}
\newcommand{\Vm}{{\bf V}}
\newcommand{\Xm}{{\bf X}}
\newcommand{\Ym}{{\bf Y}}

\newcommand{\Cc}{{\cal C}}

\newcommand{\Nc}{{\cal N}}

\newcommand{\Rc}{{\cal R}}
\newcommand{\Sc}{{\cal S}}

\newcommand{\muv}{\hbox{\boldmath$\mu$}}

\newcommand{\phiv}{\hbox{\boldmath$\phi$}}

\newcommand{\omegav}{\hbox{\boldmath$\omega$}}

\newcommand{\Gammam}{\hbox{\boldmath$\Gamma$}}
\newcommand{\Lambdam}{\hbox{\boldmath$\Lambda$}}

\newcommand{\Phim}{\hbox{\boldmath$\Phi$}}

\newcommand{\asf}{{\sf a}}

\newcommand{\Xsf}{{\sf X}}

\newcommand{\diag}{{\hbox{diag}}}

\newcommand{\trace}{{\hbox{tr}}}

\renewcommand{\Re}{{\rm Re}}
\renewcommand{\Im}{{\rm Im}}
\newcommand{\eqdef}{\stackrel{\Delta}{=}}

\newcommand{\herm}{{\sf H}}

\newcommand{\transp}{{\sf T}}

\newcommand{\SINR}{{\sf SINR}}
\newcommand{\SNR}{{\sf SNR}}

\usepackage{hyperref}
\hypersetup{
    bookmarks=true,         
    unicode=false,          
    pdftoolbar=true,        
    pdfmenubar=true,        
    pdffitwindow=false,     
    pdfstartview={FitH},    
    pdfnewwindow=true,      
    colorlinks=true,       
    linkcolor=red,          
    citecolor=green,        
    filecolor=magenta,      
    urlcolor=cyan           
}

\begin{document}

\title{Truncated Polynomial Expansion Downlink Precoders and Uplink Detectors for Massive MIMO}
\author{Andreas Benzin$^\star$, Giuseppe Caire$^\star$, \IEEEmembership{Fellow, IEEE,} Yonatan Shadmi$^\star$, and Antonia Tulino$^\dagger$, \IEEEmembership{Fellow, IEEE}
\thanks{$^\star$ Communications and Information Theory Group, Technische Universit{\"a}t Berlin (Email: \{andreas.benzin, caire\}@tu-berlin.de, yonatansha@gmail.com). $^\dagger$ Universit\'a Federico II, Napoli, Italy, and Nokia Bell-Labs, NJ, USA (Email: a.tulino@nokia-bell-labs.com).}
}

\maketitle
\begin{abstract}
In TDD reciprocity-based massive MIMO it is essential to be able to compute the downlink precoding matrix over all OFDM resource blocks 
within a small fraction of the uplink-downlink slot duration. For example, in the pioneering work by Marzetta it is assumed
that such precoding matrix can be computed within one OFDM symbol time interval. 
Early implementation of massive MIMO are limited to the simple {\em Conjugate Beamforming} (ConjBF) precoding method, because of such
computation latency limitation. However, it has been widely demonstrated by theoretical analysis and system simulation that
{\em Regularized Zero-Forcing} (RZF) precoding is  generally much more effective than ConjBF for a large but practical number of
transmit antennas. In order to recover a significant fraction of the gap between ConjBF and RZF and yet meeting the 
very strict computation latency constraints, truncated polynomial expansion (TPE) methods have been proposed.  In this paper we 
present a novel TPE method that outperforms all previously proposed methods in the general non-symmetric case of users with 
arbitrary antenna correlation. In addition, the proposed method is significantly simpler and more flexible than previously proposed methods based on
{\em deterministic equivalents} and {\em free probability} in large random matrix theory. We consider power allocation with our TPE approach, 
and show that classical system optimization problems such as {\em min sum power} and {\em max-min rate} can be easily solved. 
Furthermore, we provide a detailed computation latency analysis specifically targeted to a highly parallel FPGA hardware architecture.
Overall, our proposed TPE method used jointly with clever scheduling that arranges users in nearly mutually orthogonal groups is able to
recover almost all performance gap with respect to RZF with significantly smaller computation latency, well within the one LTE OFDM symbol
as assumed in Marzetta's work.
\end{abstract}

\begin{IEEEkeywords}
Large-scale MIMO, linear precoding, polynomial expansion, complexity reduction, uplink-downlink duality, random matrix theory.
\end{IEEEkeywords}

\newpage

\section{Introduction}  \label{sec:intro}

Massive MIMO is a very promising candidate to increase the sum-rate (bits/s/Hz) performance of future wireless networks 
by an order of magnitude compared to currently deployed cellular wireless systems \cite{mmimo_book}. 

Early works on massive MIMO proposed the use of maximum ratio transmission (MRT) in the downlink (DL) 
and maximum ratio combining (MRC)  in the uplink (UL),  since these schemes have very low precoder/detector complexity 
and can be easily distributed across different digital processing units \cite{Marzetta-TWC10}.  
It was quickly shown through theoretical analysis \cite{hoydis2011massive, Huh11} and experiments \cite{shepard2012argos}, that the regime of 
number of BS antennas per user at which MRT/MRC do not suffer from a large performance gap with respect to
their more complex counterparts, namely,  regularized zero-forcing (RZF) precoding for the DL 
and linear Minimum Mean-Square Error (MMSE) detection for the UL, is unpractically large.
As a matter of fact, in order to exploit the full potential of massive MIMO in practice, RZF precoding and MMSE detection must be implemented. 
These require the computation of matrix inverses of $K\times K$ matrices, 
where $K$ denotes the number of single-antenna users to be served simultaneously by spatial multiplexing, 
in the same time-frequency slot. Such matrix inverses must be calculated for each transmission resource block (RB) where the 
channel matrix is constant over time and frequency. 
Furthermore, even though the physical channel coherence time interval $\Delta t_c$ 
and coherence frequency $\Delta W_c$ yield a fairly number of signal dimensions $\approx \lceil \Delta t_c \times \Delta W_c\rceil$ over which the channel matrix remains constant  \cite{tse2005fundamentals}, in a practical situation where the set of $K$ active users is scheduled 
according to some dynamic scheduling algorithm operating with fast slot granularity, 
the set of users changes (due to scheduling) from slot to slot. Therefore, 
a new precoding matrix must be calculated for each scheduling slot, at a rate that may be significantly faster than the physical channel coherence. 
For concreteness, consider a massive MIMO BS serving a cell with a total of 100 connected users.
Suppose that the BS has $M = 128$ antennas and serves 
up to $K = 20$ users simultaneously, by allocating the whole bandwidth in time division, and scheduling over groups of 20 users. 
Typical values for the channel coherence time and bandwidth are  $\Delta t_c = 10$~ms and $\Delta W_c = 100$~kHz \cite{tse2005fundamentals}, yielding 
a coherence block of $\approx 1000$ signal dimensions. However, in LTE \cite{dahlman} the scheduler may operate with the granularity of a single RB, 
spanning 12 subcarriers in frequency and 14 OFDM symbols in time, for a total of $168$ signal dimensions. 
With LTE numerology, subcarriers are spaced by 15~kHz and OFDM symbols have duration 66.7~$\mu$s, such that the RB has 
duration of about 1~ms. It would be of course completely impractical to pre-calculate the precoding matrix for all possible $100 \choose 20$ scheduled users combinations. 
Therefore, the precoding matrix must be recomputed at each RB period for current the set of scheduled users, despite the fact that their physical channel matrix 
may change significantly more slowly. 

It is therefore apparent that achieving a large fraction of the precoding (resp., detection) gain 
of RFZ precoding (resp., MMSE detection) with low {\em time complexity},  such that the computation can be executed in a small fraction of the
RB period, is a key feature of any massive MIMO precoding algorithm.  We stress here the role of 
time complexity instead of the traditional Flops count for the following reason. 
With the advent of highly integrated and energy efficient Field-Programmable Gate Array (FPGA) technology, to be used specifically in the BS, 
we find more appropriate to measure algorithm complexity in terms of execution time, rather than 
in terms of Flops. This is because the number of Flops is quite immaterial, given the fact that modern FPGAs can support 
a very large number of DSP blocks for the parallel computation of additions and multiplications. 
In contrast, the suitability of algorithms to be highly parallelized plays a key role in determining whether they can meet the strict computation deadlines
necessary for massive MIMO.  

In this work we present novel a precoding method for massive MIMO DL that achieves a competitive tradeoff between performance
and time complexity. The proposed method approximates the precoding matrix by a truncated matrix polynomial expansion (TPE) 
with coefficients that can be easily pre-computed using asymptotic random matrix theory and depend only on the statistics of 
the user channels.  The TPE coefficients are obtained by first solving the dual UL problem and then using the dual UL detectors as DL precoding vectors via 
UL-DL duality \cite{Viswanath-Tse-TIT03}. This also means that the TPE solution presented here can be reused for the complexity reduction 
of the detector in the uplink, although this aspect is not explicitly addressed in this paper.  

The TPE approach 
was widely studied in the late 90s and early 2000s (see \cite{muller_verdu_2001, tulino_verdu_2004} and references therein)
to obtain efficient detectors for direct-sequence CDMA systems with long spreading codes. 
In the context of massive MIMO,  \cite{zarei_1} proposed a TPE precoding design using UL-DL duality (as in the present paper), 
assuming isotropically distributed user channel vectors, i.e., such that each vector is formed by Gaussian i.i.d. zero-mean elements, and users may differ by their pathloss. 
In this case, the dual UL TPE detector was designed to minimize the sum-MSE. Thanks to the isotropic statistics of the channel vectors, 
results from free probability led a simple method to compute the asymptotic TPE coefficients. 
A bit later, the work \cite{axel_muller_debbah_2014} considered transmit antenna correlation and measurement errors in the channel vectors. 
In \cite{axel_muller_debbah_2014} a completely symmetric scenario is considered, where all users have the same channel statistics. 
Therefore, a single set of TPE coefficients is optimized to maximize sum rate, directly for the DL. 
Further, the work \cite{hoydis_debbah_2011} considered a TPE \textit{detector} for the UL while assuming different covariance matrices for the users (and thus also 
different path losses)  and calculated the TPE coefficients that minimize the sum-MSE. 
Finally, the work \cite{kammoun2014linear} considered a non-symmetric DL situation where users can have different antenna correlations, 
and optimized the TPE coefficients directly for the DL. As such, a single set of TPE coefficients is
optimized for all users. Since optimizing the sum rate yields an intractable non-convex problem, in \cite{kammoun2014linear} a max-min approach is considered, 
where the TPE coefficients is optimized in order to maximize the minimum user rate. 
In summary, while the TPE based on large random matrix theory  approach is well-known, previous works have some shortcomings. 
On one hand, the approach in \cite{zarei_1} obtains simple and easily computable expressions, 
but must assume isotropic user channels in order to apply the machinery of free probability. On the other hand, the approach in \cite{kammoun2014linear} handle fully general
channel correlations, but needs a very complicated method to compute the asymptotic expressions (namely, a system of coupled matrix fixed-point equations that must be 
solved iteratively) and in addition can handle only one set of TPE coefficients optimized according to the max-min rate criterion. 

The TPE precoder presented in this paper overcomes the above shortcomings. Our method  handles arbitrary user channel correlations subject to a simplifying 
approximation, which is generally well satisfied when the BS is quipped with a large Uniform Linear Array (ULA). Thanks to this simplifying assumption, 
the asymptotic expressions appearing in the TPE coefficients optimization can be computed with very low complexity. 
Furthermore, by addressing the TPE optimization for the dual UL, we can optimize an individual set of TPE coefficients for each user. Also, we shall show that classical optimal power allocation problems can be easily solved for our proposed TPE precoder. As an example, we explicitly address the max-min user rate with respect to power allocation. 
Another original contribution of our work is a very detailed time complexity analysis of the proposed TPE precoder, compared with classical 
RZF precoding, whose hardware implementation has been explicitly addressed in \cite{eberli2009application,prabhu}. Our analysis specifically addresses
the time complexity in the case of an FPGA implementation and reveals that TPE precoding can be more heavily parallelized 
in digital hardware than other state-of-the art matrix inversion methods. Thus, it can accommodate significantly stricter computation deadlines arising from fast time-selective channels 
(small coherence block) or fast dynamic scheduling, as discussed above. 

The paper is organized as follows. In Section \ref{sec:problem} we define the TPE optimization problem at hand. 
In Section \ref{sec:computation} the efficient computation of the beamforming vectors is presented. 
Section \ref{sec:large-system-limit} develops the asymptotic expression for the TPE coefficient computation 
using results from large random matrix theory. Section \ref{sec:optimization} deals with optimal power allocation with TPE precoding. 
In Section \ref{sec:complexity} we present the time-complexity analysis and comparison with respect to matrix inversion. 
Section \ref{sec:results} presents some performance results, and notably a comparison with respect to the previously proposed TPE methods. 
Finally, in Section \ref{sec:conclusions} we point out some concluding remarks.

\section{Optimal TPE Precoding}  \label{sec:problem}

\subsection{Channel model and UL-DL duality}

Consider a narrow-band block-fading MU-MIMO system with a BS 
with $M$ antennas and $K$ single-antenna user equipments (UEs) with $M \gg K$. 
We model  a generic channel use of the complex baseband uplink (UL) channel at the BS receiver as
\begin{align}
	\tilde{\yv} = \tilde{\Hm} \tilde{\xv} + \tilde{\zv}, 
	\label{eq:1}
\end{align}
where $\tilde{\xv} = (\tilde{x}_1, \ldots, \tilde{x}_K)^\transp$ is the vector of user transmitted complex modulation symbols, 
the channel matrix $\tilde{\Hm} \in \CC^{M \times K}$ has columns $\tilde{\hv}_k$ corresponding to the 
user channel vectors between the transmit antenna of UE $k$ and the $M$ receive BS  antennas, and  where
$\tilde{\zv} \sim \mathcal{CN}(\zerov,  N_0 \Id_M)$ is a vector of AWGN samples at the BS baseband antenna ports.  
We assume arbitrary UL transmit power (to be optimized later), indicated by
$\EE[|\tilde x_k|^2] = \tilde{p}_k$, for $k : [1:K]$, with assigned sum power $\sum_{k=1}^K \tilde{p}_k = P$. 
For the sake of normalization of the channel coefficients,  it is convenient to define the transmit Signal-to-Noise Ratio (SNR) 
as $\SNR = P/N_0$, extract the common term $\SNR/K$, and define the normalized  power allocation coefficients $p_k$, such that the channel model in \eqref{eq:1} can be rewritten in the equivalent form
\begin{align}
	\yv =  \Hm \xv + \sqrt{\nu} \zv, 
	\label{eq:2}
\end{align}
where $\Hm = [\hv_1, \ldots, \hv_K] = \frac{1}{\sqrt M} \tilde{\Hm}$,  the $k$-th column $\hv_k$ of $\Hm$ is referred to as the (normalized) $k$-th user channel vector, 
$\beta = K/M$ is the {\em spatial loading factor} (number of users per antenna), 
$\nu = \beta/\SNR$ is the the per-component noise variance, 
$\zv \sim \Cc\Nc(\zerov, \Id_M)$,  and where we define the power allocation coefficients given by 
$\EE[|x_k|^2] = p_k \geq 0$,  for $k \in [1:K]$,  with the sum-power constraint $\sum_{k=1}^K p_k = K$. 

The {\em dual} DL  channel corresponding to (\ref{eq:2}) is given by 
\begin{equation} \label{dl}
\yv^{\rm dl} = \Hm^\herm  \Vm \sv + \sqrt{\nu} \zv^{\rm dl}, 
\end{equation} 
where $\zv^{\rm dl} \sim \Cc\Nc(\zerov, \Id_K)$, $\Vm \in \CC^{M \times K}$ is the DL precoding matrix, $\sv = (s_1, \ldots, s_K)^\transp$ is the vector of user data symbols, 
and $\yv^{\rm dl} = (y_1^{\rm dl}, \ldots, y_K^{\rm dl})^\transp$ is the corresponding vector of received observations at the UE receivers, such that 
$s_k$ and $y_k^{\rm dl}$ are the desired data symbol and the receiver observation for UE $k$, respectively. 
In practice, the two channels (\ref{eq:2}) and (\ref{dl}) may correspond to 
the UL and DL of an actual system with channel reciprocity, which can be achieved in TDD when the UL and DL slots span the same channel coherence interval 
(e.g., see \cite{Marzetta-TWC10,shepard2012argos}). Even if the the channels in (\ref{eq:2}) and (\ref{dl}) do not correspond to an actual reciprocal scenario, 
given the DL (\ref{dl}) we can always consider its dual UL in (\ref{eq:2}), whether this corresponds to an actual physical channel or not. The important point here is that 
it is much easier to run system optimization for the UL model (\ref{eq:2}) and then translate the results to the 
DL model (\ref{dl}). This translation follows from the general UL-DL duality result 
proved in \cite{Viswanath-Tse-TIT03}. This is summarized by the following theorem,   
included for the sake of completeness and since it is useful to state the result in a consistent  notation.
The details of UL-DL duality is included, for the sake of completeness, in Appendix \ref{sec:UL-DL-duality}. 

\begin{theorem}  \label{duality}
Consider the UL channel model (\ref{eq:2}) with a set of linear receiver vectors $\vv_1, \ldots, \vv_K$ with unit norm, 
such that the observation for user $k$ data stream is given by 
\begin{equation} \label{kobs}
\widehat{x}_k = \vv_k^\herm \yv = \vv_k^\herm \hv_k x_k + \sum_{j \neq k} \vv_k^\herm \hv_j x_j + \sqrt{\nu} \vv_k^\herm \zv, 
\end{equation}
yielding the Signal-to-Interference plus Noise Ratio (SINR) 
\begin{equation} \label{sinr}
\SINR_k = \frac{|\vv_k^\herm \hv_k|^2 p_k}{\vv_k^\herm \left (  \sum_{j\neq k} p_j \hv_j\hv_j^\herm  + \nu \Id_M \right ) \vv_k}.
\end{equation}
Let $\Vm = [\vv_1, \ldots, \vv_K]$ be the precoding matrix for the dual DL channel (\ref{dl}) such that 
\begin{equation} \label{kobsdl}
y^{\rm dl}_k = \hv_k^\herm \vv_k s_k + \sum_{j\neq k} \hv_k^\herm \vv_j s_j + \sqrt{\nu}  z_k^{\rm dl},
\end{equation}
with powers $q_k = \EE[|s_k|^2]$ and  resulting DL SINR 
\begin{equation} \label{dlsinr}
\SINR^{\rm dl}_k = \frac{|\hv_k^\herm \vv_k|^2 q_k}{\nu + \sum_{j\neq k} |\hv_k^\herm \vv_j|^2 q_j}.
\end{equation}
Then, there exists a set of DL powers $\{q_1, \ldots, q_K\}$ with $\sum_k q_k = \sum_k p_k$ such that 
$\SINR^{\rm dl}_k = \SINR_k$ for all $k = 1, \ldots, K$. 
\hfill $\square$
\end{theorem}

Moreover, it is very easy to compute the DL powers $\{q_k\}$ from the UL powers $\{p_k\}$, 
as shown in \cite{Viswanath-Tse-TIT03} (details are omitted for brevity).

\subsection{Optimal TPE coefficients}

In virtue of the UL-DL duality in Theorem  \ref{duality} we can just focus on the optimization of the receiver vectors for the UL model (\ref{eq:2}). 
A convenient form of the linear detector that maximizes the SINR for each UE $k$ is obtained
by first producing the {\em spatial matched filter} output 
\begin{equation}
\rv = \Hm^\herm \Hm \xv + \sqrt{\nu} \Hm^\herm \zv, 
\end{equation}
and then applying MMSE estimation of the data symbols on $\rv$.  We define the Gram matrix of the channel vectors $\Gm = \Hm^\herm \Hm \in \CC^{K \times K}$, 
such that the covariance matrix\footnote{All statistics for signal detection are conditional on the knowledge of the channel matrix $\Hm$, which is assumed to be known to the receiver.} 
of $\rv$ is given by 
\begin{equation} 
\EE[\rv \rv^\herm] = \Gm \Pm \Gm + \nu \Gm. 
\end{equation}
The MMSE linear estimator of $x_k$ from $\rv$ is given by 
\begin{align}
\hat{x}_k & = \EE[x_k \rv^\herm] \left ( \EE[\rv \rv^\herm] \right )^{-1} \rv \nonumber \\
& = p_k \gv_k^\herm \left ( \Gm \Pm \Gm + \nu \Gm \right )^{-1} \Hm^\herm \yv, 
\end{align}
where $\gv_k = \Hm^\herm \hv_k$ is the $k$-th column of $\Gm$. The resulting SINR is invariant to any (non-zero) scaling of the receiver vectors.  In particular, we choose
the SINR-maximizing receiver vector
\begin{equation} \label{mmse-v-alternative}
\vv^{\rm opt}_k = \Hm \left ( \Gm \Pm \Gm + \nu \Gm \right )^{-1} \gv_k.
\end{equation}
The resulting SINR is obtained by replacing $\vv_k = \vv_k^{\rm opt}$ in (\ref{sinr}), yielding
\begin{align} 
\SINR^{\rm opt}_k 
& = \left .  \frac{\vv_k^\herm \hv_k \hv_k^\herm \vv_k p_k}{\vv_k^\herm \left ( \Hm \Pm \Hm^\herm + \nu \Id \right ) \vv_k  - 
\vv_k^\herm \hv_k \hv_k^\herm \vv_k p_k} \right |_{\vv_k = \vv^{\rm opt}_k} \label{new-succhialemma}  \\
& = \frac{\gv_k^\herm \left ( \Gm \Pm \Gm + \nu \Gm \right )^{-1} \gv_k p_k}{1 - \gv_k^\herm \left ( \Gm \Pm \Gm + \nu \Gm \right )^{-1} \gv_k p_k}.
\label{sinr-mmse-alternative}
\end{align} 
Following the TPE approach, we wish to replace the matrix inverse in (\ref{mmse-v-alternative}) 
with a low-degree matrix polynomial.  To this purpose, we define the TPE UL receiver vector for UE $k$ in the form
\begin{align} 
\vv^{\rm tpe}_k 
& = \Hm \sum_{l=0}^J w_{k,l} ( \Pm \Gm )^l \ev_k,  \label{mmse-v-alternative-TPE}
\end{align}
where $\ev_k$ is the $k$-th column of the $K \times K$ identity matrix, and 
where $\{w_{k,l}\}$ are weights to be optimized in order to maximize the SINR of UE $k$.

Replacing $\vv_k = \vv_k^{\rm tpe}$ in (\ref{sinr}), 
after some lengthy but somehow trivial algebra (omitted for the sake of brevity),  we can write the resulting SINR in the form
\begin{equation}
\SINR^{\rm tpe}_k = \frac{\wv_k^\transp \av_k \av_k^\transp \wv_k}{\wv_k^\transp ( \Bm_k + \nu \Cm_k - \av_k \av_k^\transp) \wv_k},  \label{new-rayleigh-quotient}
\end{equation}
where the coefficients $\av_k = (a_{k,0}, \ldots, a_{k,J})^\transp \in \CC^{(J+1)\times 1}$, $\Bm_k = \{ B_{k,l,l'} : (l,l') \in [0:J] \times [0:J] \}  \in \CC^{(J+1) \times (J+1)}$, and 
$\Cm_k = \{ C_{l,l'} : (l,l') \in  [0:J] \times [0:J] \}  \in \CC^{(J+1) \times (J+1)}$ are given respectively by 
\begin{subequations} \label{aBC-coef}
\begin{align} 
a_{k,l} 
& = \bar{\hv}_k^\herm \bar{\Gammam}^l \bar{\hv}_k,   \label{new-a-coef}  \\
B_{k,l,l'} & = \bar{\hv}_k^\herm \bar{\Gammam}^{l+l'+1} \bar{\hv}_k,  \label{new-B-coef} \\
C_{k,l,l'} & = \bar{\hv}_k^\herm \bar{\Gammam}^{l+l'} \bar{\hv}_k,  \label{new-C-coef}
\end{align}
\end{subequations}
we define the scaled channel matrix incorporating the power allocation coefficients as 
$\bar{\Hm} = \Hm \Pm^{1/2}$ with $k$-th column $\bar{\hv}_k$,  and we define $\bar{\Gammam} = \bar{\Hm} \bar{\Hm}^\herm = \Hm \Pm \Hm^\herm$.

We notice that (\ref{new-rayleigh-quotient}) is a generalized Rayleigh quotient 
with positive semidefinite matrices $\av_k \av_k^\transp$ and $\Bm_k + \nu \Cm_k - \av_k \av_k^\transp$, and that the latter is strictly positive definite.\footnote{If the denominator was not strictly positive definite, one could achieve infinite capacity in the presence of white AWGN with non-zero variance and finite transmit power, which is obviously impossible.} 
By the Courant-Fischer max-min principle \cite{lancaster1985theory}, the maximum of (\ref{new-rayleigh-quotient}) is achieved by the maximum generalized 
eigenvalue of the matrix pencil $(\av_k\av_k^\transp, \Bm_k + \nu\Cm_k - \av_k \av_k^\transp)$, i.e., the maximum value $\lambda$ for which there exists a vector 
$\wv_k$ such that 
\[ \av_k \av_k^\transp \wv_k = \lambda (\Bm_k + \nu \Cm_k - \av_k \av_k^\transp) \wv_k \]
Since $(\Bm_k + \nu \Cm_k - \av_k\av_k^\transp)$ is invertible and $\av_k\av_k^\transp$ has rank 1, there is a single non-zero eigenvector 
(which therefore must be the maximum),  readily obtained by letting 
\begin{equation} \label{w-unnormalized} 
\wv_k = (\Bm_k + \nu \Cm_k - \av_k\av_k^\transp)^{-1} \av_k = \alpha_k (\Bm_k + \nu \Cm_k)^{-1} \av_k, 
\end{equation}
where the second equality follows form the matrix inversion lemma with $\alpha_k = (1 - \av_k^\transp (\Bm_k + \nu \Cm_k)^{-1} \av_k)^{-1}$. 
Since the optimal coefficients are defined up to a non-zero scaling, we choose the solution that imposes
unit-norm precoding vectors $\vv^{\rm tpe}_k$. This has also the advantage that these unit-norm vectors yield immediately the precoding vector for the DL, via UL-DL duality.  To this purpose, using (\ref{mmse-v-alternative-TPE}) we obtain
\begin{equation} \label{TPE-normalized}
\|\vv_k\|^2 = p_k^{-1} \wv_k^\transp \Cm_k \wv_k,  
\end{equation}
where $\Cm_k$ is defined in (\ref{new-C-coef}). Replacing (\ref{w-unnormalized}) into (\ref{TPE-normalized}) 
and imposing the unit-norm condition $\|\vv^{\rm tpe}_k\|^2 = 1$ we obtain the normalization coefficient
\begin{equation} \label{TPE-normalized1}
\alpha_k^* = \sqrt{\frac{p_k}{\av_k^\transp (\Bm_k + \nu \Cm_k)^{-1} \Cm_k (\Bm_k + \nu \Cm_k)^{-1} \av_k}}.
\end{equation}
The resulting normalized optimal TPE coefficients are eventually given by 
$\wv_k^* =  \alpha_k^* (\Bm_k + \nu \Cm_k)^{-1} \av_k$. 
Replacing this into (\ref{new-rayleigh-quotient}) we obtain the resulting optimal SINR with TPE detection in the form
\begin{equation} \label{opt-sinr-w}
\SINR_{k}^{\rm tpe} = \lambda_{\max}(\av_k, \Bm_k, \Cm_k) = \frac{\av_k^\transp ( \Bm_k + \nu \Cm_k)^{-1} \av_k}{1 - \av_k^\transp ( \Bm_k + \nu \Cm_k)^{-1} \av_k}.
\end{equation}

\section{Efficient Computation of the Beamforming Vectors}  \label{sec:computation}

We wish to compute $\vv^{\rm tpe}_k$ given in (\ref{mmse-v-alternative-TPE}) for all $k = 1, \ldots, K$. 
We recall Horner's rule for the efficient evaluation of a polynomial at a given argument. Let
$p(x) = p_0 + p_1 x + p_2 x^2 + \cdots + p_J x^J$. Equivalently, we can write
\begin{equation} \label{horner} 
p(x) = p_0 + x(p_1 + x(p_2 + \cdots x(p_{J-1} + p_J x)) \cdots ). 
\end{equation}
This yields the recursive computation method: let $b_J = p_J$, then, for $n = J-1, \ldots, 0$ let $b_n = p_n + b_{n+1} x$. 
It follows that $b_0 = p(x)$.  Using Horner's rule, we first focus on the computation of
the matrix polynomial $\sum_{l=0}^J w_{k,l} (\Pm\Gm)^l \ev_k$. Define $\vv_k^{(J)} = w_{k,J} \ev_k$ and let
\[ \vv^{(n)}_k = w_{k,n} \ev_k + \Pm\Gm \vv_k^{(n+1)} \]
for $n = J - 1, J-2, \ldots, 0$. It is easily shown that $\vv_k^{(0)} = \sum_{l=0}^J w_{k,l} (\Pm\Gm)^l \ev_k$. 
Now, we can replicate the computation for all UEs $k = 1, \ldots, K$ in parallel by defining
$\Wm^{(l)} = \diag(w_{1,l}, \ldots, w_{K,l})$, initializing
\[ \Vm^{(J)} = [ w_{1,J} \ev_1, \ldots, w_{K,J} \ev_K] = \Wm^{(J)} \]
and letting 
\begin{equation} \label{ziofiga} 
\Vm^{(n)} = \Wm^{(n)}  + \Pm\Gm \Vm^{(n+1)}, \;\;\; n = J-1, J-2, \ldots, 0.
\end{equation}
Finally, the matrix $\Vm^{\rm tpe} = [\vv^{\rm tpe}_1, \ldots, \vv^{\rm tpe}_K]$ of desired beamforming vectors is obtained as
\begin{equation} \label{precoding-vectors-computation}
\Vm^{\rm tpe} = \Hm \Vm^{(0)}. 
\end{equation}
In classical massive MIMO \cite{Marzetta-TWC10}, the channel matrix $\Hm$ is estimated at the BS from UL pilot symbols, exploiting TDD reciprocity.
In practice, the channel matrix $\Hm$ remains constant for a block of $T \approx \lceil \Delta t_c \times \Delta W_c\rceil$ 
channel uses in the time-frequency plane, as discussed in Section \ref{sec:intro}.
%
However, while $T$ may be quite large, due to dynamic scheduling a new precoding matrix must be calculated at each new
scheduling slot, the period of which may be significantly faster than the physical channel coherence time. 
This means that the computation of the DL precoding matrix must be accomplished in a very short time. 
For example, in \cite{Marzetta-TWC10} is it assumed that the UL slot and the DL slot are separated by 
a guard interval of one OFDM symbol (66.7~$\mu$s in LTE), necessary to calculate the DL beamforming vectors from the UL 
pilot observations.  

We remark here that there exists a method of directly calculating the precoded transmit signal vectors using TPE \cite{axel_muller_debbah_2014} which is discussed in appendix \ref{sec:dtpep}. We further show in appendix \ref{sec:dtpep} that assuming LTE system parameters TPE precoding matrix pre-calculation according to \mr{precoding-vectors-computation} and subsequent classical precoding outperforms the direct calculation variant in terms of computational latency. 

Notice that $\Vm^{(0)}$ in (\ref{precoding-vectors-computation}) is a $K \times K$ matrix, and the calculation of the 
regularized zero-forcing (RZF) precoder (DL) consist also of the multiplication of the $M \times K$ matrix $\Hm$ times
a $K \times K$ inverse matrix. Hence, the complexity comparison between the proposed TPE method 
and the standard RFZ precoder reduces to comparing the computational complexity of
the direct calculation of a $K \times K$ inverse versus the matrix polynomial $\Vm^{(0)}$. 
In Section \ref{sec:complexity} we provide a detailed comparison of the TPE computation latency 
with respect to the state-of-the-art method for direct inverse matrix computation.

\section{TPE Coefficients in the Large-System Limit}  \label{sec:large-system-limit}

The goal of this section is finding expressions for the coefficients $\av_k$, $\Bm_k$, and $\Cm_k$ appearing
in (\ref{w-unnormalized}). 
For finite $M$ and $K$, such coefficients are random and depend on the channel matrix $\bar{\Hm}$. 
If the coefficients were calculated from their finite-dimensional expressions 
(\ref{new-a-coef}), (\ref{new-B-coef}), and (\ref{new-C-coef}), the TPE method would not achieve any computational gain with respect to
direct matrix inversion. However,  following the approach widely advocated in the literature (see Section \ref{sec:intro}), 
we exploit the fact that in the large-system limit $M, K \rightarrow \infty$, with fixed ratio $\frac{K}{M} \rightarrow \beta$, the coefficients 
converge with probability 1 to deterministic quantities that can be computed only from the statistics of
the channel matrix $\bar{\Hm}$ but do not depend on its actual realization. 
Due to dynamic scheduling and the fact that the users have different channel statistics 
(in particular, different channel vector covariance matrix),
the coefficients must be computed ``on-the-fly'' for each scheduling slot. 
Hence, it is very important that the coefficients can be computed with low complexity. 
Notice that the methods based on ``deterministic equivalents'' used in \cite{shariati2014low,mueller2016linear,kammoun2014linear,lu2016low} 
involve very complicated systems  of coupled matrix fixed-point equations that must be solved iteratively, 
and whose complexity makes the TPE methods proposed in these works 
{\em completely unsuitable} for practical purposes. In contrast, here we develop  very simple expressions that require no matrix fixed-point iterations, 
and whose complexity is indeed negligible with respect to
the complexity of the beamforming vectors computation itself. 

A detailed derivation of the physical propagation channel that motivates and justifies the channel correlation model used in our paper is provided in 
Appendix \ref{sec:geometric-correlation-model}. 
The user channel vectors $\hv_k$ are mutually independent since the users are typically separated in space by several wavelengths. However, 
each user channel is a correlated Gaussian vector, with covariance matrix $\Rm_k = \EE[ \tilde{\hv}_k \tilde{\hv}_k^\herm ]$
that depends on the user-specific propagation environment.  
We develop our approach under the assumption that the BS has an ULA with a large number $M$ of antennas. 
When the BS has a Uniform Linear Array (ULA), the channel covariance matrix $\Rm_k$ is Toeplitz. 
In order to obtain simple expressions for the computation of (\ref{new-a-coef}) -- (\ref{new-C-coef})
in the asymptotic regime of large $M$ and $K$, 
we wish that all user covariance matrices have the same eigenvectors.  To this purpose, we consider the circulant approximation of 
the Toeplitz channel covariance matrices, defined as follows. Let $r^{(k)}_{m} = [\Rm_k]_{\ell,\ell-m}$, for $\ell = 0, \ldots, M-1$ and $m = 0, \ldots, \ell$, denote the $(\ell, \ell - m)$-th element of $\Rm_k$.   The circulant approximation of $\Rm_k$ is the matrix $\overset{\circ}{\Rm}_k$ whose first column has elements $[\overset{\circ}{\Rm}]_{k,m} = \overset{\circ}{r}^{(k)}_{m}$  for $m = 0, \ldots M-1$, defined by 
\begin{equation} \label{circulant-approx}
\overset{\circ}{r}^{(k)}_{m} = \left \{ \begin{array}{ll} 
r^{(k)}_0 & \mbox{for} \;\; m = 0 \\
r^{(k)}_m + r^{(k)}_{m-M} & \mbox{for} \;\; m = 1, \ldots, M-1 \end{array} \right . 
\end{equation}
Letting $\Fm$ denote the $M \times M$ unitary DFT matrix (see Appendix \ref{sec:geometric-correlation-model}), we have that $\overset{\circ}{\Rm}_k = \Fm \overset{\circ}{\Lambdam}_k \Fm^\herm$, where 
for large $M$, by Szego's Theorem (see \cite{adhikary2013joint} and references therein), $\overset{\circ}{\Lambdam}_k \rightarrow \Lambdam_k$ in the sense that 
the eigenvalue empirical cumulative distribution function of $\overset{\circ}{\Lambdam}_k$ converges pointwise almost everywhere to that of $\Lambdam_k$. 

Hence, we shall calculate the asymptotic TPE coefficients for the {\em approximated channel statistics} where the true covariance matrices  
$\Rm_k$ are replaced by their circulant approximations $\overset{\circ}{\Rm}_k$. For such approximated statistics, 
all user channel covariance matrices have exactly the same set of eigenvectors,  given by the columns of $\Fm$, 
and the error incurred in the corresponding eigenvalues vanishes as $M \rightarrow \infty$. 

%
We define the  i.i.d. Gaussian matrix $\HH = [\mathbbm{h}_1, \ldots, \mathbbm{h}_K]$ with independent elements $\sim \Cc\Nc(0,1/M)$, and  
the channel coefficient variances (including the power allocation coefficients as part of the channel) as
\begin{equation} \label{mask-coefs}
D_{m,k} = \overset{\circ}{\Lambda}^{(k)}_{m} p_k, 
\end{equation}
where $\overset{\circ}{\Lambda}^{(k)}_{m}$ is the $m$-th diagonal element of $ \overset{\circ}{\Lambdam}_k$, for $m = 0,\ldots, M-1$. 
Hence, the channel matrix approximated statistics is given by 
\begin{equation} \label{chmatrix-model}
\bar{\Hm} =  \Fm ( \HH \odot \Dm ),   
\end{equation}
where $\Dm \in \RR_+^{M \times K}$ has elements $\sqrt{D_{m,k}}$ and where $\odot$ denotes 
Hadamard product (elementwise product).

\subsection{$J$-step recursive formula} \label{sec:recursive}

Define the large-system limits of the coefficients in 
(\ref{new-a-coef}), (\ref{new-B-coef}), and (\ref{new-C-coef}) as 
\begin{subequations}  \label{aBC-asympt}
\begin{eqnarray} 
a_{k,l}^{\infty}  &=&   \lim_{M \rightarrow  \infty }
\bar{\hv}_k^\herm \bar{\Gammam}_k^l \bar{\hv}_k,   \label{new-moment-vectorasy}\\
B^{\infty} _{k,l,l'} & =& \lim_{M \rightarrow  \infty } \bar{\hv}_k^\herm \bar{\Gammam}_k^{l+l'+1} \bar{\hv}_k  \label{new-hankel-matrix-Basy} \\
C^{\infty}_{k,l,l'} & = & \lim_{M \rightarrow  \infty } \bar{\hv}_k^\herm \bar{\Gammam}_k^{l+l'} \bar{\hv}_k. \label{new-hankel-matrix-Casy}
\end{eqnarray} 
\end{subequations}
The limits (\ref{new-moment-vectorasy}) - (\ref{new-hankel-matrix-Casy}) are all special cases of the limit 
\begin{eqnarray}  \label{new-general-moment} 
\rho^{\infty} _{k,\ell} & \eqdef & \lim_{M \rightarrow  \infty }  \bar{\hv}_k^\herm \bar{\Gammam}^\ell \bar{\hv}_k,
\end{eqnarray} 
for $\ell \in \ZZ_+$.  The main difficulty of computing $\rho^{\infty} _{k,\ell}$ consists of the fact that the channel vector
$\bar{\hv}_k$ appears also inside the matrix $\bar{\Gammam}$, thus creating a dependency between the terms of the 
quadratic form $\bar{\hv}_k^\herm \bar{\Gammam}^\ell \bar{\hv}_k$. We circumvent this problem by establishing a $J$-step 
recursive formula that yields
$\rho^{\infty} _{k,0}, \ldots, \rho^{\infty} _{k,J}$ in terms of the sequence $\gamma^\infty_{k,0}, \ldots, \gamma^\infty_{k,J}$, with
\begin{eqnarray}  \label{general-moment} 
\gamma^{\infty} _{k,\ell} & \eqdef & \lim_{M \rightarrow  \infty }  \bar{\hv}_k^\herm \bar{\Gammam}_k^\ell \bar{\hv}_k,
\end{eqnarray} 
where $\bar{\Gammam}_k = \sum_{j \neq k} \bar{\hv}_j \bar{\hv}_j^\herm = \bar{\Hm}_k  \bar{\Hm}_k^\herm$, and where
$\bar{\Hm}_k \eqdef [ \bar{\hv}_1, \ldots, \bar{\hv}_{k-1}, \bar{\hv}_{k+1}, \ldots, \bar{\hv}_K]$
does not contain $\bar{\hv}_k$. 
A closed-form expression (up to definite integrals)  for $\gamma^{\infty} _{k,\ell}$ is provided later in Section \ref{sec:original-coefs}.
The relation between $\rho^{\infty} _{k,\ell}$ and $\gamma^{\infty} _{k,\ell}$ is obtained by taking the limit for $M \rightarrow \infty$
in the following finite-dimensional result:

\begin{lemma} \label{rho-gamma-lemma}
For any finite $M, K$ and given matrix $\bar{\Hm}$, the quadratic forms $\rho_{k,\ell} =  \bar{\hv}_k^\herm \bar{\Gammam}^\ell \bar{\hv}_k$
and $\gamma_{k,\ell} =  \bar{\hv}_k^\herm \bar{\Gammam}^\ell_k \bar{\hv}_k$ are related by 
\begin{equation} \label{recursion} 
\rho_{k,\ell} = \gamma_{k,\ell} + \sum_{i=1}^{\ell} \gamma_{k,\ell - i} \rho_{k,i-1}, 
\end{equation} 
\end{lemma}

{\em Proof:}
We write
\begin{eqnarray}
\rho_{k,\ell}  & = & \bar{\hv}_k^\herm \bar{\Gammam}^{\ell-1} \left ( \bar{\Gamma}_k + \bar{\hv}_k\bar{\hv}_k^\herm \right ) \bar{\hv}_k \nonumber \\
& = & \bar{\hv}_k^\herm \bar{\Gammam}^{\ell-1} \bar{\Gammam}_k \bar{\hv}_k + \| \bar{\hv}_k\|^2 \rho_{k,\ell-1} \nonumber \\
& = & \bar{\hv}_k^\herm \bar{\Gammam}^{\ell-1} \bar{\Gammam}_k \bar{\hv}_k + \gamma_{k,0} \rho_{k,\ell-1}  \label{recursion1}
\end{eqnarray}
where we used the fact that $\gamma_{k,0} = \rho_{k,0} = \| \bar{\hv}_k\|^2$. 
Let's focus on the first term in (\ref{recursion1}). We have
\begin{eqnarray}
\bar{\hv}_k^\herm \bar{\Gammam}^{\ell-1} \bar{\Gammam}_k \bar{\hv}_k  & = & \bar{\hv}_k^\herm \bar{\Gammam}^{\ell-2} \left ( \bar{\Gamma}_k + \bar{\hv}_k\bar{\hv}_k^\herm \right )   \bar{\Gammam}_k \bar{\hv}_k \nonumber \\
& = & \bar{\hv}_k^\herm \bar{\Gammam}^{\ell-2} \bar{\Gammam}^2_k \bar{\hv}_k + \bar{\hv}_k^\herm \bar{\Gammam}_k \bar{\hv}_k \bar{\hv}_k^\herm \bar{\Gammam}^{\ell-2} \bar{\hv}_k \nonumber \\
& = & \bar{\hv}_k^\herm \bar{\Gammam}^{\ell-2} \bar{\Gammam}^2_k \bar{\hv}_k + \gamma_{k,1}  \rho_{k,\ell-2}   \label{recursion2}
\end{eqnarray}
We can apply again same decomposition to the first term in (\ref{recursion2}), obtaining
\begin{equation} 
\bar{\hv}_k^\herm \bar{\Gammam}^{\ell-2} \bar{\Gammam}^2_k \bar{\hv}_k = \bar{\hv}_k^\herm \bar{\Gammam}^{\ell-3} \bar{\Gammam}^3_k \bar{\hv}_k + \gamma_{k,2}  \rho_{k,\ell-3}. 
\end{equation}
This  is repeated such that at the $\ell$-th step the residual term reduces to  $\gamma_{k,\ell}$. Putting all terms together, we arrive at (\ref{recursion}).
\hfill $\square$

\subsection{Computation of $\gamma_{k,\ell}^\infty$}  \label{sec:original-coefs}

We recall here the definition of variance profile \cite[Def. 2.16]{tulino2004random}:
\begin{definition}
For a random matrix of the type (\ref{chmatrix-model}) with uniformly bounded elements $D_{m,k}$, 
the variance profile  $v^M : [0,1) \times [0,1) \rightarrow \RR_+$ is the piecewise constant function
\begin{equation} \label{varianceprofile}
v^M(x,y) = D_{m,k}, \;\;\; \frac{m-1}{M} \leq x < \frac{m}{M}, \;\;\; \frac{k-1}{K} \leq y < \frac{k}{K}.
\end{equation}
Whenever $v^M(x,y)$ converges uniformly to a limiting bounded measurable function $v(x,y)$, this limit is referred to as the
{\em asymptotic variance profile} of $\bar{\Hm}$. \hfill $\lozenge$
\end{definition}
An expression for $\gamma_{k,\ell}^\infty$ is established by the following result, which follows as a rather immediate 
corollary of \cite[Th. 2.56]{tulino2004random}:

\begin{theorem}  \label{theorem-antonia1}
Consider an $M \times K$ matrix $\bar{\Hm}$ defined as in (\ref{chmatrix-model}). 
Assume that, as $M, K \rightarrow \infty$, 
the asymptotic variance profile $v(x,y)$ exists. The limit of the quadratic form $\gamma^\infty_{k,\ell}$ defined in (\ref{general-moment})
converges almost surely to the function
\begin{eqnarray} \label{gamma-infinity}
\gamma^{\infty} _{\ell}(y) = \int_0^1 \xi _{\ell}(x) v(x,y) dx, \;\;\;\; y \in [0,1)
\end{eqnarray}
calculated at $\frac{k-1}{K} \leq y < \frac{k}{K}$,  where the function $\xi_\ell(x)$ is given by 
$\xi_0(x) = 1$ for $\ell = 0$, and for $\ell \geq 1$ through the recursion
\begin{eqnarray} \label{recursion-xi}
\xi_\ell(x)  &=& \beta \xi_{\ell-1}(x) \int_0^1 v(x,y) dy \nonumber \\
& & + \beta \sum_{j=1}^{\ell-1}  \xi_{j-1}(x) \int_0^1 v(x,y)
\sum_{i=1}^{\ell-j} \sum_{\footnotesize{n^{(i)}_1+\cdots+n^{(i)}_i = \ell-j}} \;\;
\prod_{\mu=1}^i \left [ \int_0^1 v(x',y) \xi_{n^{(i)}_\mu-1}(x') dx' \right ]  dy.
\end{eqnarray}
Furthermore, the asymptotic moments of $\bar{\Hm}$ are given by
\begin{eqnarray} \label{moments-infinity}
\lim_{M \rightarrow \infty}  \frac{1}{M} \trace \left ( \left ( \bar{\Hm} \bar{\Hm}^\herm \right )^\ell \right )  
= \int_0^1 \xi_{\ell}(x) dx 
\end{eqnarray}

\hfill $\square$
\end{theorem}

\begin{remark} The notation $\sum_{i=1}^{\ell-j} \sum_{\footnotesize{n^{(i)}_1+\cdots+n^{(i)}_i = \ell-j}}$ indicate the sum 
over all possible {\em ordered} integer partitions of the integer $\ell - j$,. For $1 \leq i \leq \ell - j$, 
the ordered integer partitions of order $i$ of the integer $\ell - j$ are the distinct ordered sets $\{n^{(i)}_1,\ldots, n^{(i)}_i\}$ 
of {\em strictly positive} integers such that $n^{(i)}_1 + \cdots + n^{(i)}_i = \ell - j$. \hfill $\lozenge$
\end{remark}


{\bf Finite dimensional approximation:}
In practice, in order to compute $\gamma^{\infty}_{k,\ell}$ defined in (\ref{general-moment}) 
we use a discretized form of Theorem \ref{theorem-antonia1}. Namely, for sufficiently large $M,K$,  
we can use the approximation
$$ v(x,y) \approx D_{m,k},  \quad  x = \frac{m}{M},  \quad   y = \frac{k}{K} $$  
for $m = 0, \ldots, M-1$ and $k = 0, \ldots, K-1$, such that (\ref{gamma-infinity}) yields
\begin{eqnarray} \label{discrete-general-moment}
\gamma^{\infty} _{k,\ell} \approx  \frac{1}{M} \sum_{m=0}^{M-1} \xi _{\ell}\left (\frac{m}{M} \right ) D_{m,k},  
 \end{eqnarray} 
where for $m = 0,\ldots, M-1$ we have
\begin{eqnarray} \label{discrete-recursion}
\xi_{\ell}\left (\frac{m}{M} \right )  
& = & \beta \xi_{\ell-1}\left (\frac{m}{M} \right )  \frac{1}{K} \sum_{k=0}^{K-1} D_{m,k} \nonumber \\
& & + \beta \sum_{j=1}^{\ell-1}  \xi_{j-1}\left (\frac{m}{M} \right ) \frac{1}{K} \sum_{k=0}^{K-1} D_{m,k}
\sum_{i=1}^{\ell-j} \sum_{\footnotesize{n^{(i)}_1+\cdots+n^{(i)}_i = \ell-j}} \;\;
\prod_{\mu=1}^i \left [ \frac{1}{M} \sum_{m'=0}^{M-1} D_{m',k} \xi_{n^{(i)}_\mu-1}\left (\frac{m'}{M}\right )  \right ]  
\end{eqnarray}

\begin{example}
For the sake of clarity, let's write explicitly the recursive calculation in Theorem \ref{theorem-antonia1} in the case $\ell = 3$. 
We start by calculating $\xi_1(x)$. From (\ref{recursion-xi}), this is given by 
\begin{equation} \label{xi1}
\xi_1(x) = \beta \xi_0(x) \int_0^1 v(x,y) dy = \beta \int_0^1 v(x,y) dy.
\end{equation}
Next, we use again (\ref{recursion-xi}) to calculate $\xi_2(x)$. This is given by the sum of two terms for $j = 1,2$ in the outer sum. We have
\begin{eqnarray} 
\xi_2(x) & = & \beta \int_0^1 v(x,y) \left ( \int_0^1 v(x',y) dx' \right ) dy + \beta \xi_1(x) \int_0^1 v(x,y) dy
\end{eqnarray}
where we used the fact that for $j = 1$ and $\ell = 2$ there is only one unique integer partition of the integer $\ell - j = 1$ formed 
by $n^{(1)}_1 = 1$. 

Finally, $\xi_3(x)$ is given by the sum of three terms for $j = 1,2,3$ in the outer sum. We have
\begin{eqnarray}
\xi_3(x) & = & \beta \int_0^1 v(x,y) \left [ \int_0^1 v(x',y) \xi_1(x') dx' + \left ( \int_0^1 v(x',y) dx' \right )^2 \right ] dy \nonumber \\
& & + \beta \xi_1(x) \int_0^1 v(x,y) \left ( \int_0^1 v(x',y) dx' \right ) dy \nonumber \\
& & + \beta \xi_2(x) \int_0^1 v(x,y) dy
\end{eqnarray}
where we used the fact that for $j = 1$ and $\ell = 3$ there are two unique integer partitions of the integer $\ell - j = 2$, namely
the partition of order 1, $n^{(1)}_1 = 2$, and the partition of order 2, $n^{(2)}_1 = 1, n^{(2)}_2 = 1$.   

Particularizing the above formulas for the case of $v(x,y) = 1$, i.e., the case where $\bar{\Hm}$ is i.i.d., we find
\begin{subequations}
\begin{eqnarray}
\xi_0(x) & = & 1 \label{xi0-iid}\\
\xi_1(x) & = & \beta \label{xi1-iid}\\
\xi_2(x) & = & \beta + \beta^2 \label{xi2-iid}\\
\xi_3(x) & = & \beta + 3 \beta^2 + \beta^3 \label{xi3-iid}
\end{eqnarray}
\end{subequations}
As expected, for this i.i.d. example the obtained expressions coincide with the first moments of the Marcenko-Pastur distribution, 
the general expression of which is given by \cite{tulino2004random}
\begin{equation} \label{mp-moments} 
\EE[ \xi_{\ell}(\Xsf) ]  = \frac{1}{\ell} \sum_{i=1}^\ell {\ell \choose i}   {\ell \choose i-1}  \beta^i 
\end{equation}
\hfill $\lozenge$
\end{example}

\section{Power Allocation with TPE Precoding}  \label{sec:optimization}

In this section we consider canonical power allocation problems applied to the case of linear TPE precoding. 
Using UL-DL duality it is convenient to formulate the power allocation for the dual UL channel, and then translate the result into the corresponding 
DL power allocation and precoding vectors. In general, since the optimal TPE  coefficients $\{\wv_k\}$ in (\ref{w-unnormalized}) 
depend on the power allocation through the large-system limit coefficients $\av_k, \Bm_k$ and $\Cm_k$, 
which in turn depend on the power allocation matrix $\Pm$ via the weighting matrix $\Dm$ in (\ref{mask-coefs}) - (\ref{chmatrix-model}), 
it follows that the optimization of the powers yields also the corresponding optimal TPE coefficients. 

\subsection{Conventional power control}  \label{sec:conventional-pc}

In this case, the power of each UE is allocated in order to undo the effect of its corresponding 
large-scale pathloss (including a distance-dependent attenuation and 
possibly log-normal shadowing). The large-scale pathloss coefficient (also referred to as ``channel strength'') is given by 
$A_k = \frac{1}{M} \trace(\Rm_k)$.  Hence, the conventional power control allocates the UL powers according to
\begin{equation} 
p_k = \frac{1}{A_k} \left ( \frac{1}{K} \sum_{j=1}^K \frac{1}{A_j} \right )^{-1}, 
\end{equation}
where it is easily checked that the constraint $\sum_{k=1}^K p_k = K$ is satisfied. 

\subsection{Minimum power for assigned target SINRs}

The classical min-power problem can be written as \cite{yates1995framework} 
\begin{subequations}  \label{min-power}
\begin{eqnarray}
\mbox{minimize} & & \sum_{k=1}^K p_k \\
\mbox{subject to} & &  \SINR_k^{\rm tpe} \geq \underline{\SINR}_k, \;\;\; \forall \; k,
\end{eqnarray}
\end{subequations}
where $\{\underline{\SINR}_k\}$ are given individual SINRs guarantees.  
Problem (\ref{min-power}) can be solved by the 
iterative Yates-Foschini-Milianic algorithm \cite{yates1995framework,foschini1993simple} thanks to the following result: 

\begin{lemma} \label{standard-interference} 
The function of the power allocation vector $\pv = (p_1, \ldots, p_K)^\transp$ given by
\begin{equation} 
I_k(\pv) = \frac{p_k \underline{\SINR}_k}{ \lambda_{\max}(\av_k,\Bm_k,\Cm_k)}, 
\end{equation}
where $\lambda_{\max}(\av_k,\Bm_k,\Cm_k)$ is given in (\ref{opt-sinr-w})  is a {\em standard interference function} in the sense of \cite{yates1995framework}.
\end{lemma}

{\em Proof.}
Let $\vv^{\rm tpe}_k$ denote the TPE unit-norm receiving vector  for user $k$ with optimized weights for given power allocation $\pv$. 
Then, by construction we have (see (\ref{new-succhialemma}))
\[  \lambda_{\max}(\av_k,\Bm_k,\Cm_k) = p_k \frac{| (\vv^{\rm tpe}_k)^\herm \hv_k|^2}{\sum_{j \neq k}{| (\vv^{\rm tpe}_k)^\herm \hv_j |^2} p_j + \nu}  = p_k \mu_k(\pv) \]
with $\mu_k(\pv) \eqdef  \lambda_{\max}(\av_k,\Bm_k,\Cm_k) / p_k$, so that we can write $I_k(\pv) = \underline{\SINR}_k/\mu_k(\pv)$. 
Next, we shall check that $I_k(\pv)$ satisfies the definition of standard interference function given in \cite{yates1995framework}, i.e., it satisfies the following conditions:
\[ 
\left \{ \begin{array}{rl}
I_k(\zerov) \geq 0 & \mbox{positivity} \\
I_k(\pv') \geq I_k(\pv) \;\; \mbox{if} \;\; \pv' \geq \pv \;\;  & \mbox{monotonicity} \\
\alpha I_k(\pv)  > I_k(\alpha\pv) \;\; \forall \; \alpha > 1 & \mbox{scalability}
\end{array} \right . \]
Positivity and monotonicity are immediate. 
In order to prove scalability, we notice that 
scaling all powers by a factor $\alpha > 1$ is equivalent to replace the noise variance 
$\nu$ with $\nu/\alpha$. Let $\vv^{(\alpha)}_k$ denote the optimized TPE receiver 
for noise variance $\nu/\alpha$ and powers $\pv$.  Since by construction this maximizes the SINR, we have
\begin{eqnarray}
\mu_k(\alpha \pv) = \frac{\underline{\SINR}_k}{I_k(\alpha \pv)} & = & 
\frac{| (\vv^{(\alpha)}_k)^\herm \hv_k|^2}{\sum_{j \neq k}{| (\vv^{(\alpha)}_k)^\herm \hv_j |^2} p_j + \frac{\nu}{\alpha}} \nonumber \\
& \geq & 
\frac{| (\vv^{(1)}_k)^\herm \hv_k|^2}{\sum_{j \neq k}{| (\vv^{(1)}_k)^\herm \hv_j |^2} p_j + \frac{\nu}{\alpha}} \nonumber \\
& \geq & 
\frac{| (\vv^{(1)}_k)^\herm \hv_k|^2}{\sum_{j \neq k}{| (\vv^{(1)}_k)^\herm \hv_j |^2} p_j + \nu} \nonumber \\
& = & \frac{\underline{\SINR}_k}{I_k(\pv)} > \frac{\underline{\SINR}_k}{\alpha I_k(\pv)}   \label{figata}
\end{eqnarray}
where the strict inequality in (\ref{figata}) follows from the fact that $\alpha > 1$. \hfill $\square$

As a consequence of Lemma \ref{standard-interference} and of the theory developed in \cite{yates1995framework}, we have that if 
(\ref{min-power}) is feasible, then the component-wise minimum power vector supporting the individual SINR guarantees 
$\{\underline{\SINR}_k\}$ is obtained by the unique fixed point of the recursion
\begin{equation} \label{yates-recursion}
p_k^{(t + 1)} = I_k(\pv^{(t)}), \;\;\; t = 0,1,2,\ldots, 
\end{equation} 
with $\pv^{(0)} = \epsilon \onev$ for some small $\epsilon > 0$. In contrast, if (\ref{min-power}) is unfeasible, then some components of the vector $\pv^{(t)}$ will grow unbounded as $t \rightarrow \infty$. 
In the large system limit, we shall use the asymptotic expressions (\ref{aBC-asympt}) in the power control recursion (\ref{yates-recursion}), such that 
the resulting power allocation coefficients $\pv$ do not depend on the instantaneous realization of the channel matrix, but only on its statistics. 
This is in line with the fact that in massive MIMO we have a channel hardening effect, such that it is not worthwhile to allocate the power according to the 
fluctuations of the small-scale fading \cite{Marzetta-TWC10,Huh11,hoydis2013massive}. 

\subsection{Max-Min SINR}  \label{sec:max-min}

A classical ``hard-fairness'' network optimization consists of maximizing the minimum user rate, which in our context is equivalent to maximizing the minimum SINR (e.g., see \cite{kammoun2014linear}).
The optimization problem can be written as
\begin{subequations} \label{max-min-rate1}
\begin{eqnarray} 
\mbox{maximize} & & \xi \nonumber \\
\mbox{subject to} & &  \SINR^{\rm tpe}_k \geq \xi, \;\;\; \forall \; k \label{max-min-constraint} \\
  & & \sum_{k=1}^K p_k = K, \;\; p_k \geq 0.
\end{eqnarray}
\end{subequations}
This can be solved by a two level-approach, where the outer optimization 
consists of a simple line search (e.g., by bi-section) over the parameter $\xi$, 
and in the inner problem (for fixed $\xi$) consists of solving (\ref{min-power}) with the common SINR guarantee $\underline{\SINR}_k = \xi$ for all $k$. 
In the bisection line search, we shall increase or decrease the value of $\xi$ until the resulting minimum sum power to support target SINRs equal to $\xi$
satisfies the constraint $\sum_{k=1}^K p_k = K$ within a prescribed tolerance. In order to perform the bisection search we need a range for $\xi$. 
A trivial lower bound is $\xi = 0$, while an immediate upper bound is given by $\xi_{\max} = \SNR \max_k \trace(\Rm_k)$. Hence, 
the bisection search can be initialized with search interval $[0, \xi_{\max}]$. 

\subsection{Sum-Rate maximization}  \label{sec:max-sumrate}

The maximization of a weighted sum of instantaneous rates is at the core of a whole class of optimal scheduling schemes
solving the general Network Utility Maximization (NUM) problem
\begin{subequations}
\begin{eqnarray}  \label{num}
\mbox{maximize} & & U(\overline{\Rm}) \\
\mbox{subject to} & & \overline{\Rm} \in \Rc(\SNR)
\end{eqnarray}
\end{subequations}
where $\Rc(\SNR)$ denotes the {\em ergodic} achievable rate region of the system, and where 
$U(\cdot)$ is a general concave non-decreasing function of the ergodic rates. In fact, it can be shown (e.g., see \cite{shirani2010mimo} and references therein) that 
the NUM problem (\ref{num}) can be solved to any degree of accuracy by the following ``on-line'' scheduling algorithm
based on the idea of ``virtual queues''. The virtual queue algorithm works as follows.  Each user is associated to a virtual queue with backlog 
$Q_k(t)$ at slot time $t$. These are initialized to some suitable non-negative value $Q_k(0)$ at time 0. Then, 
at each slot time $t = 0,1,2, \ldots$, the algorithm calculates the instantaneous rates as the solution of 
the maximum weighted sum rate (MWSR): 
\begin{subequations}
\begin{eqnarray} \label{mwsr}
\mbox{maximize} & & \sum_{k=1}^K Q_k(t) \log(1 + \SINR_k(t)) \\
\mbox{subject to} & & \sum_{k=1}^K p_k = K, \;\;\; p_k \geq 0 \;\; \forall \; k
\end{eqnarray}
\end{subequations}
where $\SINR_k(t)$ denotes the SINR of user $k$ for the fading state $\Hm = \Hm(t)$ at time $t$,  and then updates the virtual queue backlogs 
according to
\begin{equation} \label{virtual-queue-update}
Q_k(t + 1) = \left [ Q_k(t) - R_k(t) + B_k(t) \right ]_+,
\end{equation} 
where $R_k(t) = \log(1 + \SINR_k(t))$ denotes the instantaneous rate of user $k$ at slot $t$ resulting from 
the solution of (\ref{mwsr}) and where $\{B_k(t) : k = 1, \ldots, K\}$ is the solution of the optimization problem
\begin{subequations}
\begin{eqnarray} \label{arrival-process}
\mbox{maximize} & & V U(\bv) -  \sum_{k=1}^K Q_k(t) b_k  \\
\mbox{subject to} & & b_k \in [0, B_{\max}]  \;\; \forall \; k. 
\end{eqnarray}
\end{subequations}
Standard results in stochastic optimization show that when the channel state process $\Hm(t)$ is 
stationary and ergodic, by choosing $B_{\max}$ large enough such that $\Rc(\SNR) \subseteq [0, B_{\max}]^K$, 
the time-averaged rates
\[ \overline{R}_k = \lim_{t \rightarrow \infty} \frac{1}{t} \sum_{\tau = 0}^{t - 1} R_k(\tau) \]
generated by this algorithm approach the optimal solution of (\ref{num}) by less than a per-component additive gap
that decreases as $O(1/V)$. Hence, by making the parameter $V$ sufficiently large, we can approach the optimum point for any 
network utility function. For example, a popular and widely used class of network utility functions is given by the so-called 
$\alpha$-fairness functions given in \cite{mo2000fair} which include $U(\overline{\Rm}) = \sum_k \log\overline{R}_k$ (proportional fairness)
and $U(\overline{\Rm}) = \min_k \overline{R}_k$ (max-min throughput fairness). The solution of (\ref{arrival-process}) is usually very simple since
this is a convex problem and for most common network utility functions its solution can be given in closed from. 
Instead, the solution of the MWSR problem (\ref{mwsr}) is generally difficult. In particular, for multiuser MIMO channels with 
linear receivers (UL) or linear precoding (DL), the MWSR problem is non-convex and generally there exist no known method to
solve it exactly and efficiently. In our case, the problem is further complicated by the fact that we restrict the 
receivers/precoders to be the class of TPE defined before. 

Hence, we shall address the MWSR problem (\ref{mwsr}) by powerful heuristics. First of all, since we focus on massive MIMO, 
it is reasonable to assume that the SINR achieved by any user with optimized TPE coefficients is significantly larger than 1. 
Hence, we shall approximate $\log(1 + \SINR_k) \approx \log \SINR_k$. 
For a given set of weights $\{Q_k\}$, we shall maximize $\sum_{k=1}^M Q_k \log \SINR_k$ by fixing the TPE coefficients, and optimizing with respect to the 
power allocation vector $\pv$, and then re-calculating the TPE coefficients for the new power allocation vector according to 
(\ref{w-unnormalized}). The optimization of $\pv_k$ for fixed $\{\wv_k\}$ is obtained using the 
{\em Geometric Programming} approach (e.g., see \cite{chiang2007power}). In particular, for fixed receiver vectors $\{\vv_k\}$ define
$\phi_{k,j} = |\vv_k^\herm \hv_j|^2$ such that
the SINR of user $k$ from (\ref{new-succhialemma}) can be written as
\begin{equation} 
\SINR_k = \frac{\phi_{k,k} p_k}{\sum_{j\neq k} \phi_{k,j} p_j + \nu}.  
\end{equation}
Defining the non-negative auxiliary variables $T_k = \frac{1}{\SINR_k}$, the MWSR problem can be approximated as
\begin{subequations}
\begin{eqnarray} \label{mwsr1}
\mbox{minimize} & & \prod_{k=1}^K  T_k^{Q_k} \\
\mbox{subject to} & & \frac{\nu + \sum_{j\neq k} \phi_{k,j} p_j}{\phi_{k,k} p_k} \leq T_k \;\;\; \forall \; k \\
& & p_k \geq 0, \;\; T_k \geq 0, \;\; \forall \; k
\end{eqnarray}
\end{subequations}
where (\ref{mwsr1}) is recognized to be a Geometric Program in posynomial form, that can be very efficiently solved by standard methods. 
The power allocation vector $\pv$ resulting from the solution of (\ref{mwsr1}) is then used to recompute the TPE coefficients 
and the process is iterated till convergence. It can be shown that this scheme converges to a local maximum of the 
approximated weighted sum rate $\sum_k Q_k \log \SINR_k$. 
 
\section{FPGA Implementation Complexity and Latency Analysis} \label{sec:complexity}

As already remarked in Section \ref{sec:computation}, in massive MIMO the computation latency issue is particularly stringent in the DL, 
since the precoding matrix (or the precoded signal vector block) must be computed immediately after the current channel matrix $\Hm$ is available to the BS \cite{Marzetta-TWC10}. 
Because of this extremely low latency constraint, we focus our complexity analysis on the DL case. 
Furthermore, we focus specifically on the time-complexity (i.e., computation latency) in the presence of a modern 
FPGA hardware architecture, with a high degree of parallelization. 

The analysis presented here shows that the computational latency is strongly related to the degree of parallelizability of the algorithms. 
The most competitive (in terms of numerical performance) classical matrix inversion algorithms have intrinsic dependencies in their procedures 
which result in a computational latency that can't be lowered further even if these algorithms are implemented in highly parallel architectures. 
In contrast, we shall show that the TPE algorithm (\ref{ziofiga}) can be much better parallelized than state-of-the-art matrix inversion algorithms, so that 
it can deliver significantly faster precoding matrix computation, provided that a sufficient amount of parallel hardware resource is used. 

{\bf Implementation environment assumptions.}
Let $N_{\m{CM}}$, $N_{\m{CA}}$, $N_{\m{M}}$, and $N_{\m{A}}$, denote the number of clock cycles required to perform one complex multiplication (CM), 
complex addition (CA), real multiplication (M), and real addition (A), respectively. In a modern FPGA, 
dedicated hard 2-input-operand real multipliers and real adders are available and can be implemented from standard libraries \cite{xilinx_dsp,altera_dsp}. We 
quantify the required amount of hardware resource in terms of DSP blocks (resource unit D), 
where $1\times \m D$ consists of one hard real multiplier and one hard real adder. The most time and resource consuming signal processing 
task in the RZF and TPE algorithms is the calculation of inner products. The inner product of $S$-dimensional vectors $\ve a$ and $\ve b$ 
via $\ve a^\herm\ve b = \sum_{s=1}^S a_s^* b_s = c$ with $\ve a, \ve b \in \C^S, c \in \C$ requires
\begin{align}\label{m:dot_prod_lat}
	N_{\m{CM}} + \log_2(S) N_{\m{CA}} = N_{\m{M}} + N_{\m{A}} + \log_2(S) N_{\m{A}}
\end{align}
clock cycles while using 
\begin{align}
	S \times \m{CM} + (S-1) \times \m{CA} = 4S \times \m M + 2S \times \m A + 2(S-1) \times \m A = 4S \times \m M + (4S-2) \times \m A \approx 4S \times \m D
\end{align}
DSP blocks on the assumed FPGA like implementation environment.  The $\log_2(S)$ factor in \mr{m:dot_prod_lat} 
is due to the usage of a binary adder tree in order to perform the summation of the $S$  products $a_s^* b_s$ with two-operand adder units. 
In this example, we employ $S$ complex multiplier (CM) blocks, such that all $S$ complex multiplications can be performed 
in parallel in $N_{\m{CM}}$ clock cycles. 

\subsection{Classical matrix inversion algorithms}

In order to make a fair and realistic comparison, we consider the matrix inversion algorithm delivering the best numerical fixed-point stability 
and which is most likely to be the algorithm of choice for real-time implementation in the massive MIMO application. In particular, 
\cite{eberli2009application} has shown that matrix inversion through QR decomposition using Givens rotations (QRG) 
or Householder reflections (QRH) delivers best BER performance in the high SNR regime for MMSE MIMO detectors. 
All other competitor methods for matrix inversion (QR decomp. using Gram-Schmidt, LR decomp., LDL decomp., Cramer's rule, Rank-1 updates) 
are numerically inferior in fixed point environments \cite{eberli2009application}. 
Here we do not consider complexity-reduced QR decomposition algorithms for massive MIMO precoding as presented in \cite{prabhu}, 
since these algorithms have inferior BER performance with respect to the exact QRH or QRG algorithms. 
In addition, these reduced complexity schemes rely on the massive MIMO regime to take advantage of properties of the channel Gram matrix 
(like the diagonal dominance) and thus their performance depends on the specific channel conditions. Finally, in the context of parallelizability, 
reduced complexity schemes as in \cite{prabhu} do not bring any sizeable gain. 
As a matter of fact, for a fair comparison under our architectural assumptions, the algorithm with higher parallelizability 
should be chosen.  As the analysis in \cite{cosnard1986complexity} suggests, QRH can be better parallelized than QRG. 
The underlying reason is that while QRH can zero out any number of entries in a column-vector in one operation, 
QRG can only zero out one scalar entry per operation. 

\begin{figure}[t]
	\centering	 \includegraphics[width=6cm]{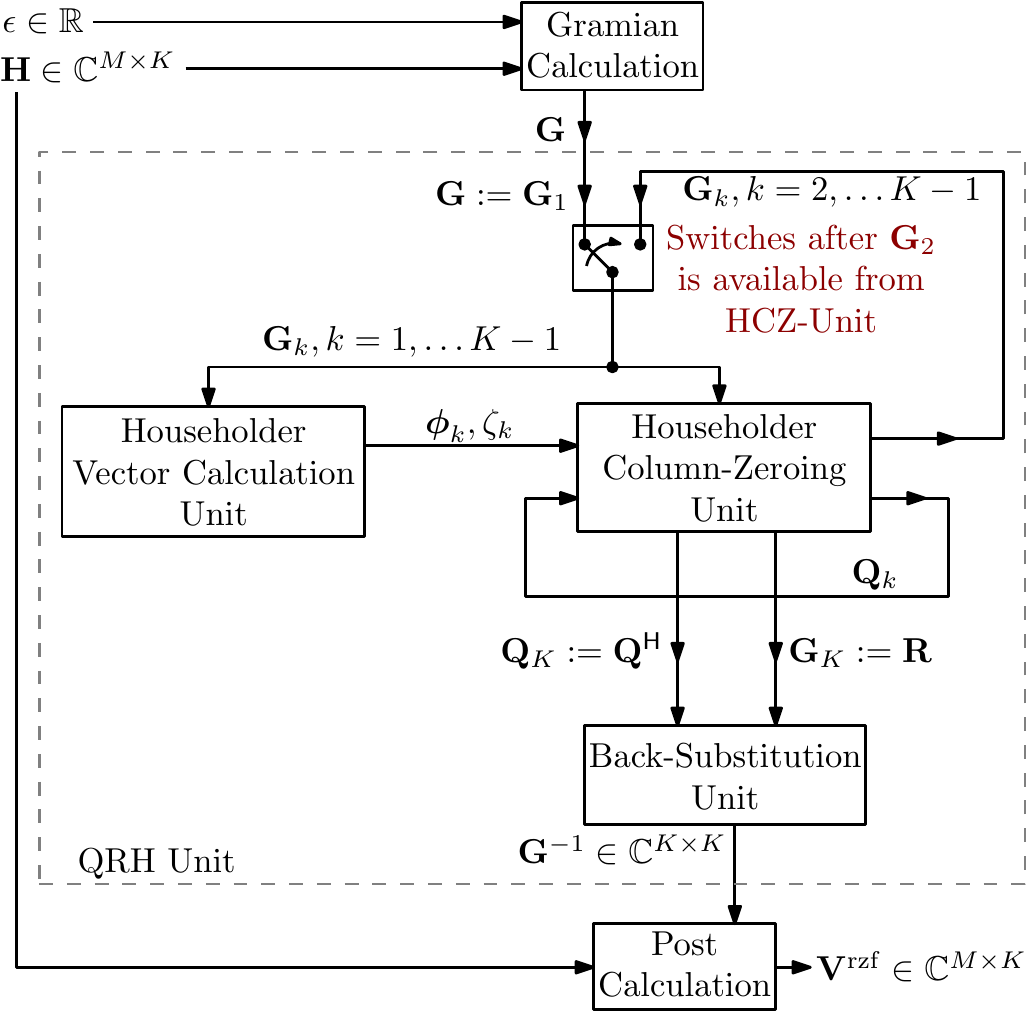}
	\caption{RZF precoding matrix calculation employing matrix inversion via QR decomposition with Householder reflections.}
	\label{fig:classical_precoder_overview}
\end{figure}

{\bf Maximally parallelized architecture of RZF precoding computation.} 
An overview of the maximally parallelized algorithm which performs the calculation of the RZF precoding matrix $ \ma V^{\rm rzf} = \ma H( \ma H^\herm \ma H + \epsilon \ma I_K)^{-1} = \ma H \ma G^{-1}$, 
using the aforementioned QR decomposition with Householder transformations, is provided in Fig.~\ref{fig:classical_precoder_overview}. 
The input of the algorithm is the channel matrix $\Hm$ and the regularization factor $\epsilon$. 
The output is the precoding matrix.  First the {\em regularized} Gram matrix $\ma G =\ma H^\herm \ma H + \epsilon \ma I_K$ has to be calculated, 
which is done in the {\em Gramian Calculation} (GC) unit. Then, the result is fed into the QRH unit which computes the matrix inverse using Householder transformations 
and back-substitution. This is accomplished by first feeding $\Gm$  into the {\em Householder Vector Calculation} unit (HVC), which calculates the first normal vector 
$\phiv_1$ of the hyperplane across which the first column vector of the Gramian is reflected and the first normalization 
scalar $\zeta_1 = 2/(\phiv_1^\herm \phiv_1)$. These two are fed into the {\em Householder Column-Zeroing} (HCZ) unit, which performs the $K$ reflections 
of the $K$ column vectors in the $k$-th iteration Gramian $\ma G_k$. The resulting matrix $\ma G_{k+1}$ is fed again into the HVC unit to calculate the second normal vector and this loop goes on until $k=K$, where $\ma G_K$ is triangular now. While the HVC unit is running, the HCZ unit is re-used to calculate the orthogonal matrix $\ma Q$ from the $k$-th iteration $\ma Q_k$s, which are the linear transformation matrices 
performing the $k$-th iteration Householder reflection. After the last iteration, the QR decomposition of $\ma G$ is obtained and this is then used in the back-substitution (BKS) unit  to calculate the inverse of the regularized channel Gramian. 
Eventually, in order to obtain the RZF precoding matrix $\ma V^{\rm rzf} = \ma H\ma G^{-1}$, the matrix inverse must be multiplied by the channel matrix $\ma H$. 
This is done in the {\em Post-Calculation} (PC) unit. 

\subsection{RZF and TPE latency and resource complexity}

This section provides the results of the computation latency and hardware resource analysis for the different algorithm 
units. The detailed algorithmic description for the discussed QRH unit and the sub-units is provided in appendix \ref{sec:qrh_details}. 
The latency $L$ and required resources $\chi$ for the calculation of one Householder vector $\phiv_k$ and normalization 
scalar $\zeta_k$ (HVC unit), the column zeroing operations for one Gramian matrix 
$\ma G_k$ (HCZ unit) and the back substitution (BKS unit) in a maximally parallelized fashion are given by 
\begin{align}
	L_{\m{HVC}} &= 2N_{\m{CCM}} + (4+\lceil\log_2(K-2)\rceil)N_{\m{A}} + N_{\m M} + 2N_{\m{RD}} + N_{\m S}\text{, }\\
	L_{\m{HCZ}} &= 2N_{\m{CM}} + (1+\lceil\log_2(K-k+1)\rceil)  N_{\m{CA}} \text{, }\\
	L_{\m{BKS}} &= K(N_{\m{CD}} + N_{\m{CM}} + N_{\m{CA}} )\text{, }
\end{align}
and by
\begin{align}
\chi_{\m{QRH}} \approx 4(K^2+3K) \times \m  D,
\end{align}
respectively, where CCM, CD, RD, and S denote complex conjugate multiplication, 
complex division, real division and the real square-root operation, and  $N_{\m{CCM}}$, $N_{\m{CD}}$, $N_{\m{RD}}$, and $N_{\m S}$, 
denote the corresponding required number of clock cycles. The detailed derivation of the resource estimate $\chi_{\m{QRH}}$ is provided in appendix \ref{sec:qrh_details}. 

In order to cancel the $K-1$ column vectors in the channel Gramian, the HVC and HCZ units have to be executed $K-1$ times to obtain the triangular form. 
This is another intrinsic algorithm dependency of the QR decomposition which cannot be parallelized.  During all iterations except for the first one, the HCZ unit computes the Householder 
reflector aggregator matrix $\ma Q_{k+1}$ at the same time when the HVC unit calculates 
the next Householder normal vector, i.e. $\phiv_{k+1}$. This is possible because $L_\m{HVC} > L_\m{HCZ}$ holds. 
After the last iteration the HCZ unit has to be called one last time to calculate the final $\ma Q_K$. Thus,  the HCZ unit is called $K$ times and the HVC unit is called $K-1$ times. 
Since the BKS unit has to wait for the $K$ iterations before it can start to operate, its latency simply adds to the overall latency of the QRH unit. 
The latency of the QRH unit is therefore given by
\begin{align}\label{m:qrh_latency}
	L_\m{QRH} = (K-1)L_{\m{HVC}} + L_{\m{BKS}} +\left( K(2 N_{\m{CM}} + N_{\m{CA}})+ N_{\m{CA}} + N_{\m{CA}}\sum\limits_{k=1}^{K-1} \lceil\log_2(K-k+1)\rceil \right).
\end{align}
This is the number of clock cycles needed to calculate the inverse of a $K\times K$ matrix assuming a maximally parallelized FPGA implementation of the QR decomposition via Householder reflections and back-substitution. 

In order to enable a meaningful comparison with TPE implemented via the {\em TPE Recursion} (TR) \mr{ziofiga},  
we introduce an architecture whose hardware resource is parameterized by a parallelization index $U$. 
Specifically, we assume to devote
\begin{align}
	\chi_\m{TR}(U) &= U \left ( K^2 \times \m{CM} + K(K-1) \times \m{CA} \right ) = 4UK^2 \times \m{D} \quad \m{with}\;  U = 2^n, n=1,2,\ldots ,\log_2(K),  
\end{align}
resources for the TR unit.   For $U = K$ we obtain the maximally parallelized implementation.  
Since the TR unit mainly calculates matrix-matrix products, we assume an implementation with $UK$ dot-product processors, 
where each dot-product processor  can compute a dot product of two $K$ dimensional vectors. Then, the TR unit has computational latency
\begin{align}\label{m:tr_lat}
			L_\m{TR}(U) &= N_\m{CM} + N_\m{CA} + (J-1) \left( \left( N_\m{CM} + \log_2(K) N_\m{CA} \right) + \left( \frac{K}{U}-1 \right)  +  N_\m{CA} \right) 
\end{align}
clock cycles. Since the main contribution to the overall TR computation consists of matrix-matrix products, 
which can be heavily parallelized, the TR unit can greatly benefit from pipelining when resources are constrained. 
For $U=K$ the parallelization is maximum, latency is minimal, but no pipelining effects can be exploited. 
This is evidenced by the fact that  $K/U-1=0$ in \mr{m:tr_lat}. 
In this case, the computation time is equal to the initial latency to fill the pipe and when the first result exits the pipe this is already the final result. 
Decreasing resources (i.e., choosing $U < K$) doesn't has much dramatic effect on the latency, since
 the TR unit can generate an intermediate result every clock cycle by exploitation of pipelining. 

Since both the GC and the PC units have to run in a sequential fashion before and after the QRH or TR units respectively, 
we can assume that the resources from the QRH or the TR unit are reused and thus no additional resources are required.  In this case the latency of the GC unit will be
\begin{align}\label{m:gc_pipe}
	L_\m{GC}(U) = N_\m{A} + N_\m{CM} + \begin{cases}\lceil\log_2(M)\rceil N_\m{CA}  + \left(\frac{M}{U}-1\right)  & \text{if } K^2\ge M\\ 
                                    \left(\lceil\log_2(K^2)\rceil + \lceil\log_2\left(\frac{M}{K^2}\right)\rceil\right) N_\m{CA}  + \left(\frac{M}{U}-1\right)\left(1+\frac{M}{K^2}\right) & \text{if } K^2 < M \end{cases}
\end{align}
and for the PC unit the latency will be
\begin{align}\label{m:pc_lat_pipe}
	L_\m{PC}(U) =  \underbrace{(N_\m{CM} + \lceil \log_2(K)\rceil N_\m{CA} )}_{\text{initial latency}} + \underbrace{\left( \frac{M}{U}-1 \right)}_{\text{remaining cycles when pipe full}}. 
\end{align}

\subsubsection{Latency amplification of precoding matrix calculation}

We compare the latency of the computation of $\ma V^{\rm tpe}$ with respect to $\ma V^{\rm rzf}$ via QRH as a function of the provided number of hardware resources measured 
in units of DSP blocks. Furthermore, we consider only the range of resources (in DSP units) such that the QRH unit can run at it's full parallelized implementation (as discussed before). For the DSP unit's behavior we're going to assume $N_\m{A} = 1$ and $N_\m{M} = 1$, i.e. one real multiplication takes one clock cycle and the same for one real addition. Furthermore we assume that the division and square-root operations require $N_\m{RD} = 4$, and $N_\m{S} = 4$ clock cycles, 
which is a reasonable choice considering that these operations are implemented using fast division methods (e.g. the Newton-Raphson method \cite{koren}).

\begin{figure}[t]
	\centering	 \includegraphics[width=12cm]{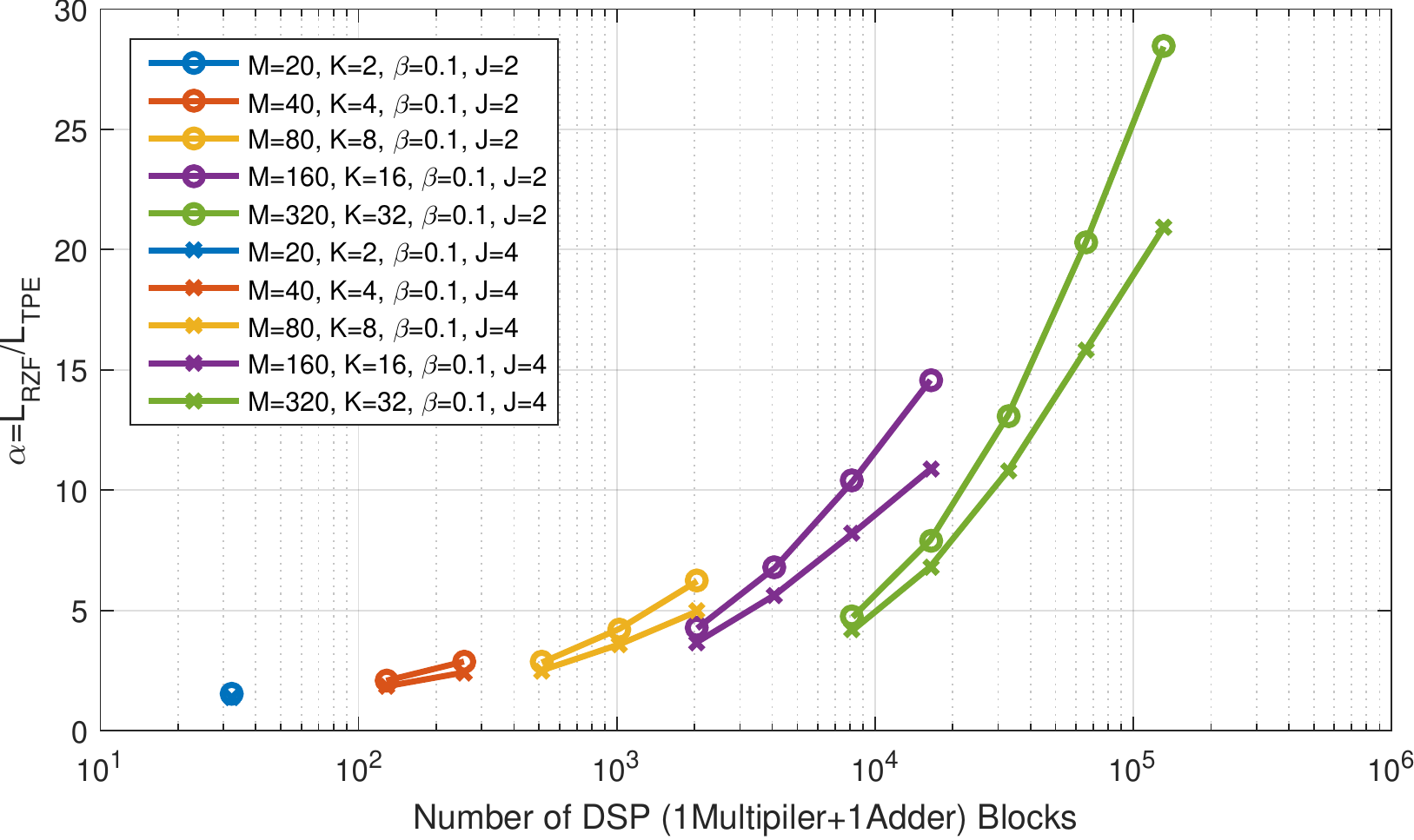}
	\caption{Latency amplification of precoding matrix calculation using QRH in relation to TPE for $\beta = 0.1$ 
	and different values of BS antennas $M$, mobile stations $K$ and matrix polynomial degree $J$ as a function of the provided number of hardware resources}
	\label{fig:rzf_tpe_vs_resources_beta_8}
\end{figure}

Given the channel matrix $\ma H$, the TPE polynomial weights $\{w_{k,l} : k = 1,\ldots, K, \; l=0,1,\ldots , J\}$, and the RZF regularization factor $\epsilon$, 
the total number of clock cycles required to compute a precoding matrix via QRH decomposition is given by 
\begin{align}\label{m:rzf_lat}
	L_\m{RZF}(U) =  L_\m{GC}(U) + L_\m{QRH}+ L_\m{PC}(U)
\end{align}
and corresponding latency for TPE via the TR algorithm is given by 
\begin{align}
	L_\m{TPE}(U) =  L_\m{GC}(U) + L_\m{TR}(U) + L_\m{PC}(U).
\end{align}
We define the ratio
\begin{align}
	\alpha(U) = \frac{L_\m{RZF}(U) }{L_\m{TPE}(U)}, 
\end{align}
as the {\em latency amplification} of RZF with respect to TPE.  
Fig.~\ref{fig:rzf_tpe_vs_resources_beta_8} shows the latency amplification $\alpha$ for different values of $M$, $K$ and $J$, 
as a function  of the number of hardware DSP blocks. 

Our results show that the TPE precoding matrix calculation runs faster than the RZF algorithm for each parameter combination considered. 
Furthermore, we can observe that, for all $U \geq 2$, the latency amplification increases with the number of antennas and users. 
For $U=2$ the QRH unit runs at full parallelization, since $\chi_\m{TR}(U=2) \approx 8K^2 \m{D} \approx \chi_\m{QRH} $ holds. This regime corresponds to the 
leftmost point of each line in Fig.~\ref{fig:rzf_tpe_vs_resources_beta_8}. For each set of parameters, the latency gain of TPE versus RZF increases with $U$ since the TR unit can benefit 
from the higher degree of parallelization while the RZF unit does not benefit in equal manner due to its intrinsic algorithmic dependencies. 
As it can be clearly seen in \mr{m:rzf_lat}, the RZF algorithm can only benefit from higher amount of resources during the calculation of the Gramian 
and the post calculation,  since these matrix-matrix products can be further parallelized. 

\subsubsection{Computation latency for practical situations}
Given our latency expressions it is also possible to estimate the required number of clock cycles to compute a precoding matrix using TPE. 
For example, with $M = 160$, $\beta = 0.1$, $J=4$ and $\chi = 4096 \times \m D$, TPE needs $\approx 130$ clock cycles. 
Assuming a digital processing clock of e.g. $f_d = \un{300}{MHz}$ (which is not unreasonable due to the large amount of pipelining employed in the implementation assumption) 
and an OFDMA system with $B = 100$ resource blocks (which corresponds to e.g. \un{20}{MHz} LTE channel bandwidth \cite{dahlman})
the calculation of the $100$ precoding matrices will take 
\begin{align}
	\frac{B}{f_d} \cdot L_\m{TPE}\left(M=160, K=16, U=\frac{4096}{4\cdot 16^2}, J=4\right) =  \frac{100}{\un{300}{MHz}}\cdot 130 \approx \un{43.3}{\mu s}.
\end{align}
This should be compared with the LTE OFDM symbol duration of $\approx \un{66.7}{\mu s}$, such that the whole precoder can be calculated
in less than one OFDM symbol.  In order to obtain the computation time for the same set of parameters and conditions in the case of RZF precoding, 
we just have to multiply the calculated time with the corresponding value of the latency amplification in Fig.~\ref{fig:rzf_tpe_vs_resources_beta_8}, 
which is about 5.6. This results in a computation time of $\approx \un{243}{\mu s}$, i.e., requiring more than 3 OFDM symbols.  


\section{Results} \label{sec:results}

\begin{figure}[ht]
\centerline{\includegraphics[width=6cm]{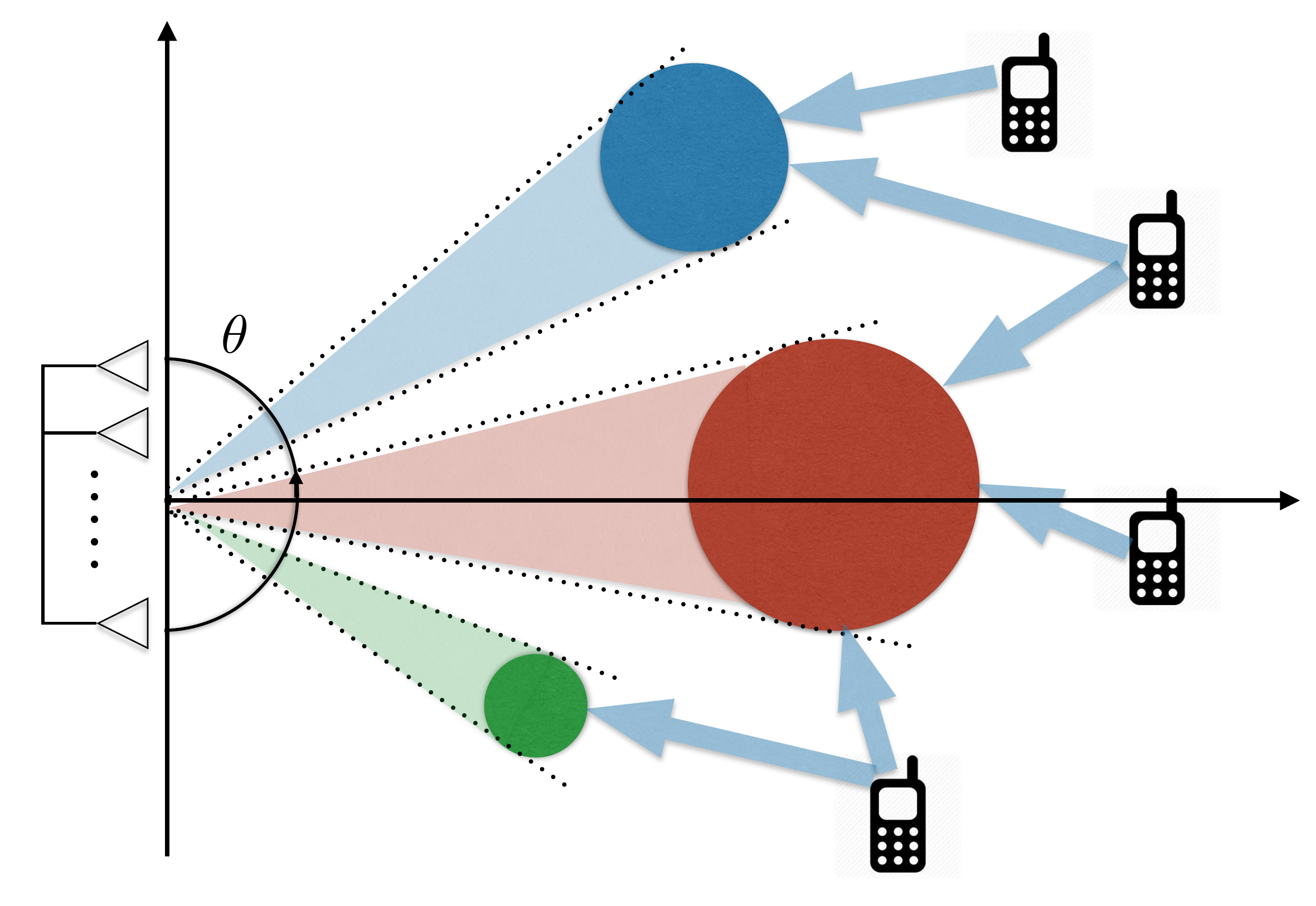}}
\centering
\caption{A qualitative representation of the scattering environment considered in our simulations, consistently with
several recent works dealing with correlated channels in massive MIMO (e.g., \cite{gao2015spatially,rao2014distributed}).}
\label{clusters}
\end{figure}

In this section we provide numerical examples illustrating the performance of the proposed TPE method. 
For simplicity, we considered a single-cell setting where the BS is equipped with an ULA with $M = 160$ antenna elements 
and antenna spacing $d = \lambda/2$ ($\lambda$ denoting the carrier wavelength), and serves
$K = 16$ users on each scheduling slot, corresponding to a loading factor $\beta = K/M = 0.1$. 
The propagation environment is formed by $S$ scattering clusters. 
Each user connects to the BS through a (non-empty) subset of 
such scattering clusters (see \cite{gao2015spatially,rao2014distributed}), as qualitatively depicted in Fig.~\ref{clusters}. 
We let $\Sc_k \subseteq [1:S]$ denote the set of scattering clusters coupled with user $k$. 
With reference to the channel model presented in Appendix \ref{sec:geometric-correlation-model}, 
each cluster $s \in [1:S]$ is characterized by an angular scattering function $\rho^{(s)}(\theta)$. The channel coefficients belonging to different scattering clusters are uncorrelated. 
Therefore, the scattering function $\rho_k(\theta)$ of user $k$ is given by $\rho_k(\theta) = \sum_{s \in \Sc_k} \rho^{(s)}(\theta)$. 
In particular,  we fixed $\rho^{(s)}(\theta)$ to be constant over the support $[\theta_s - \Delta_s/2: \theta_s + \Delta_s/2]$ and zero elsewhere, 
where  $\theta_s$ and $\Delta_s$ denote the $s$-th cluster center Angle-of-Arrival (AoA) and Angular Spread (AS), respectively. 
The cluster angular supports are contained in the interval $[-\pi/3,\pi/3]$ (i.e., a sector of width 120$^{\circ}$ centered around the array 
broadside direction $\theta = 0$, see Fig.~\ref{clusters}).   The cluster scattering functions are normalized such that $\int \rho^{(s)}(\theta) d\theta = 1/S$,  
yielding  $A_k = |\Sc_k|/S$.\footnote{In these simulations we have assumed normalized pathloss equal to 1 in order to put more in evidence the  differences between TPE methods and the role of different channel correlations, rather than the effect of the pathloss. 
Of course, the theory is general and situations with pathloss imbalance can be easily studied and reveal trends analogous to those in the results of this section. 
However,  we do not include such results here for the sake of space limitation and since they are not very informative.}

Notice that, although we have used the assumption of common eigenvectors in our large-system TPE coefficient derivation in Section \ref{sec:large-system-limit}, 
the results of this section are obtained by generating channels with the finite-dimensional covariances given by (\ref{channel-cov}), 
for which the (finite $M$) eigenvectors are not exactly the same for all users since the finite-dimensional Toeplitz corelation matrices do not exactly coincide with their 
circulant approximation. This small mismatch between the assumed channel statistics in the large-system limits
and the actual channel statistics is taken into account in our finite-dimensional simulations. 

\begin{figure}[ht]
\centerline{\includegraphics[width=8cm]{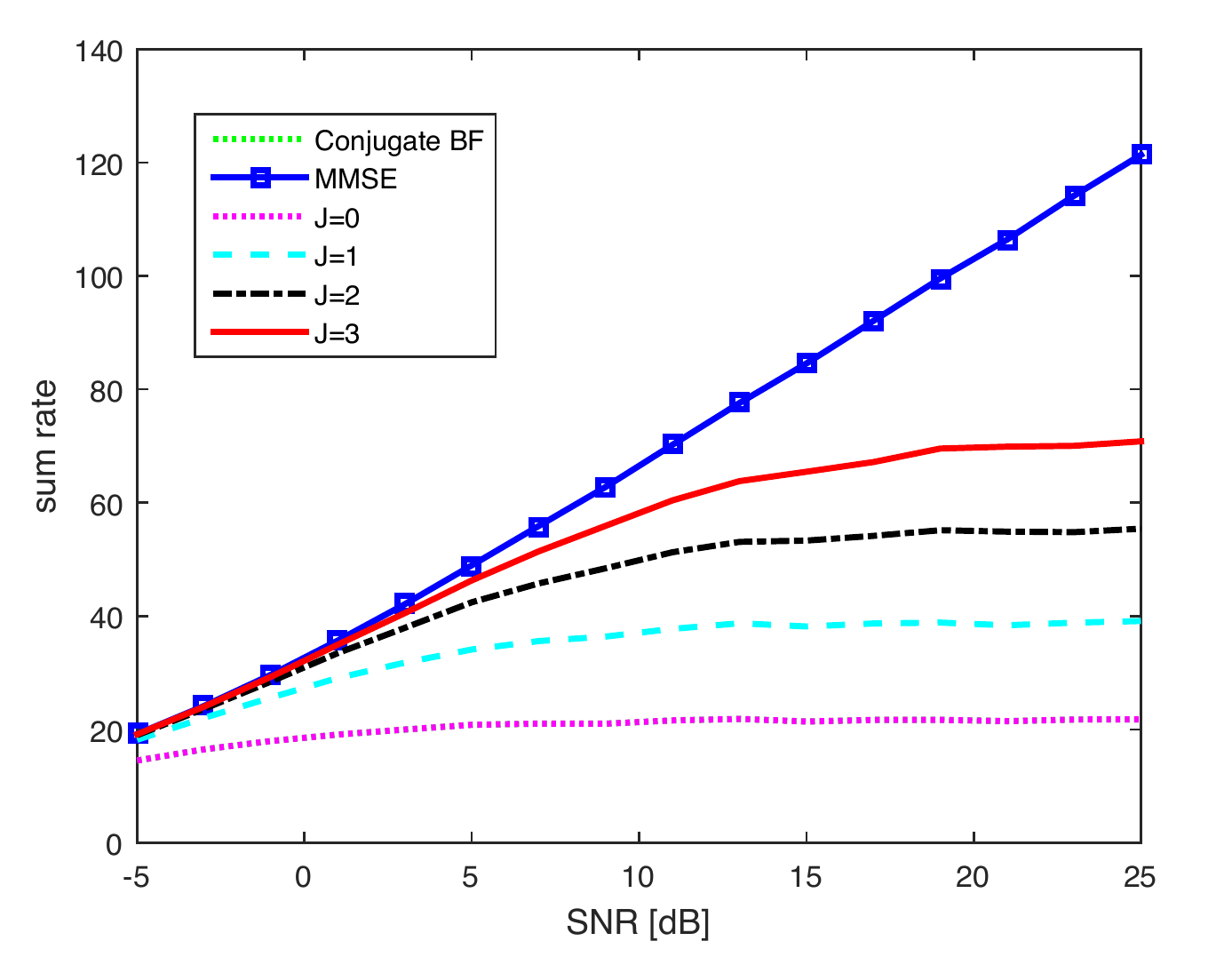} \hspace{1cm} \includegraphics[width=8cm]{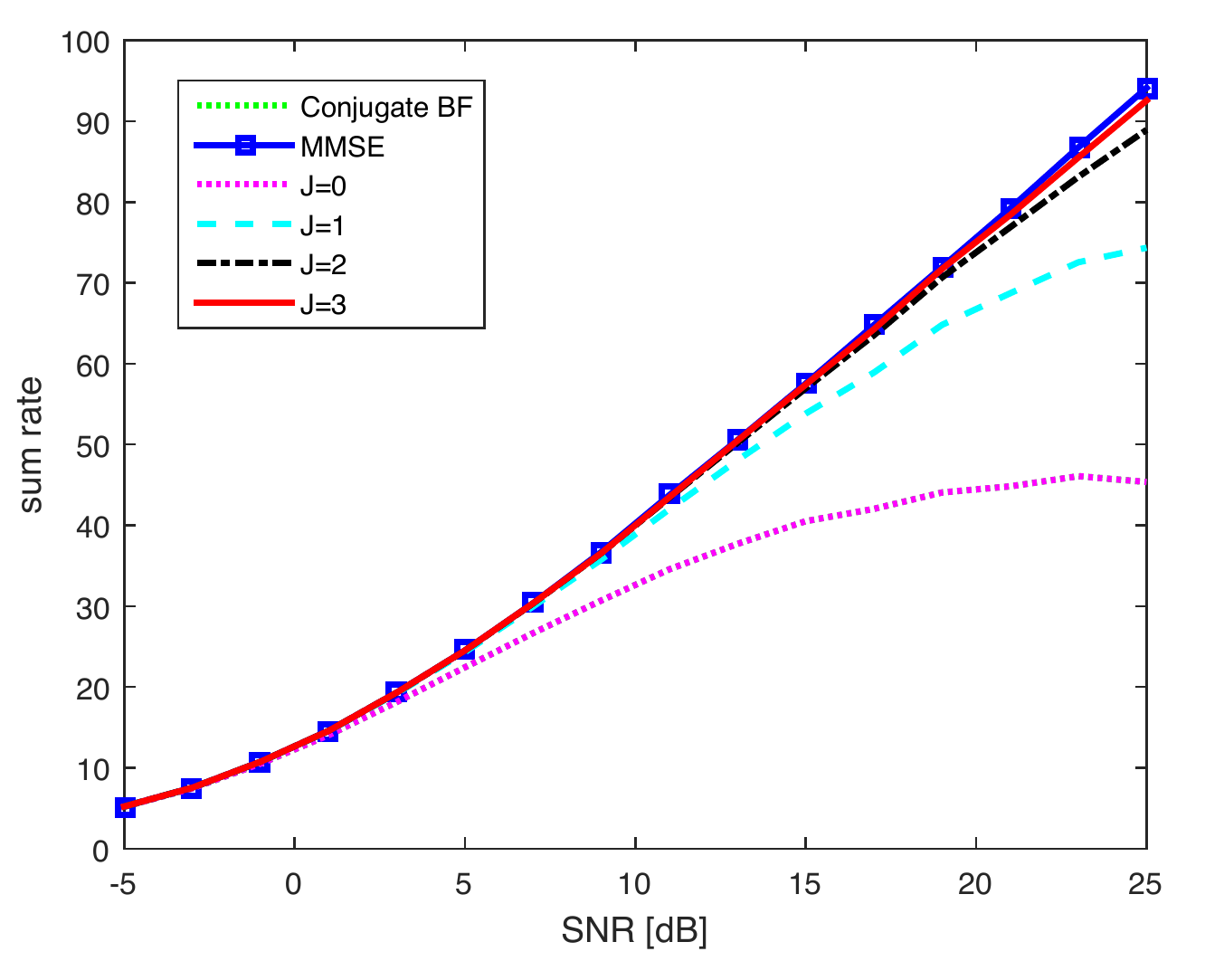}}
\centering
\caption{Effects of user channel correlation on the TPE performance. 
(left) Ergodic sum rate vs. SNR for different TPE order and users with the same antenna correlation. (right) 
Ergodic sum rate vs. SNR for different TPE order and users with different antenna correlations and quasi-orthogonal groups.}
\label{Jplotex1}
\end{figure}

First, we consider the effect of the polynomial order $J$ in the TPE precoding performance.
Fig.~\ref{Jplotex1} shows the ergodic sum rate, defined as  $\sum_k \EE[ \log(1 + \SINR^{\rm tpe}_k)]$, 
where expectation is with respect to the small-scale fading,  as a function of the system $\SNR$ (see model (\ref{dl}), 
for different polynomial expansion orders $J = 0,\ldots, 3$ and two different scattering geometries.  
In the first geometry (Fig.~\ref{Jplotex1}, left), we consider a single cluster for all the users, such that all users have the same covariance matrix. 
The cluster support has center AoA $0^{\circ}$ and AS $30^{\circ}$. In the second geometry (Fig.~\ref{Jplotex1}, right), 
we considered 8 scattering clusters with non-overlapping angular support. 
Each scattering function has AS $180^{\circ}/11$ and the center AoAs of the 8 groups are equally spaced 
such that the total support included in the interval $[-60^{\circ},60^{\circ}]$. Each user is coupled with a single cluster, and 
each cluster is connected with two users. 
The performances of the Conjugate BF \cite{Marzetta-TWC10} and of the ideal MMSE receiver/precoder 
$\vv^{\rm opt}$ (\ref{mmse-v-alternative}) are also shown for comparison.\footnote{Notice that, in this case, the precoder $\Vm^{\rm opt}$ 
 obtained by the dual MMSE corresponds to the RZF precoder with an optimal regularization factor.}
Notice that in both cases the TPE with $J = 0$ coincides exactly with
the Conjugate BF.  These results are obtained for uniform (dual UL) power allocation, i.e., $p_k = 1$ for all $K$.  
Fig.~\ref{Jplotex1} points out the fact that when the users have significantly different covariance matrices, and 
their covariance eigenspaces are clustered in small subsets of mutually quasi-orthogonal groups (as in the case of  (Fig.~\ref{Jplotex1}, right)), the TPE precoder with
low degree $J = 1$ or $J = 2$ is sufficient to recover the gap between Conjugate BF and the optimal linear precoder (MMSE/RZF).
In fact, in the case where the users are exactly mutually orthogonal, it is obvious to see that Conjugate BF (equivalently, the TPE precoder with $J = 0$) coincides 
with the optimal linear precoder. In contrast, when many user have the same correlation, i.e., they are associated to the same scattering cluster, 
even a TPE degree as high as $J = 3$ is not enough to close the gap with respect to the optimal linear precoder. Nevertheless, even in this case, 
the performance gain of $J = 2$ is significant (more than 100\% spectral efficiency gain with respect to the Conjugate BF at moderate to high SNR). 

Next, we compared our method with the other most recent TPE methods for DL precoding given 
in \cite{zarei2013low} and in  \cite{kammoun2014linear}. For completeness, these schemes and their assumptions are reviewed and presented in the
unified notation of this paper in Appendices \ref{sec:mueller} and \ref{sec:debbah}, respectively. 
The comparison is made for $J = 3$, and includes the performance of the 
Conjugate BF and MMSE receiver/precoder for the sake of comparison. 
This time, we choose  a mixed geometry  with $S = 5$ clusters with equal AS given by $\Delta_s = \pi/6$ and center AoAs 
$[-30.62^{\circ},-17.56^{\circ},-16.69^{\circ},7.5^{\circ},11.92^{\circ}]$. The association of users to clusters is defined Table \ref{table1}.

\begin{table}
\begin{center}
\begin{tabular}{|cc||c|c|c|c|c|c|c|c|c|c|c|c|c|c|c|c|} 
\hline
    & users     & 1 & 2 & 3& 4 & 5 & 6 & 7 & 8 & 9 & 10 & 11 & 12 & 13 & 14 & 15 & 16 \\ 
clusters &&  & &  & & & & & & & & & & & & & \\ \hline \hline
1            &&*  & &  & *& *&* & & & & &* & & &* & & * \\ \hline
2            &&  & & * & & *& & & & *& &* & *& & & * & * \\ \hline
3            && * &* & * & & & & & *&* &* &* & *& *& *& & \\ \hline
4            && * & *& * &* & *& & & & & &* & *& *& *& *& *\\ \hline
5            &&  & * & *  &*  &*  & &*  & & & * & * & * & * & * & & * \\ \hline
\end{tabular}
\end{center}
\caption{Cluster-users association for the geometry of Fig.~\ref{Jplotex2}. User $k$ is connected to cluster $s$ if there is 
an asterisk ``*'' in the $k$-th position of the $s$-th row.}
\label{table1}
\end{table}

\begin{figure}[ht]
\centerline{\includegraphics[width=8cm]{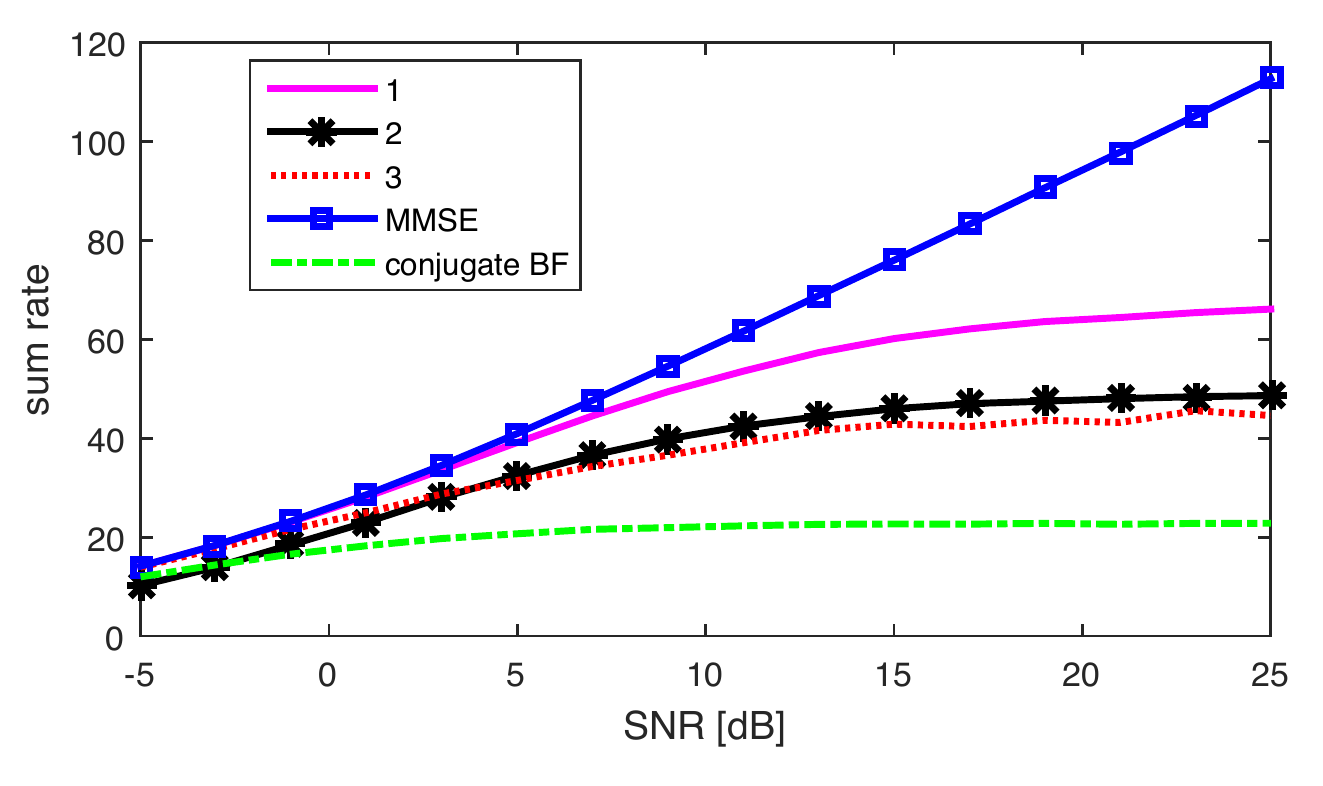} \hspace{1cm} \includegraphics[width=8cm]{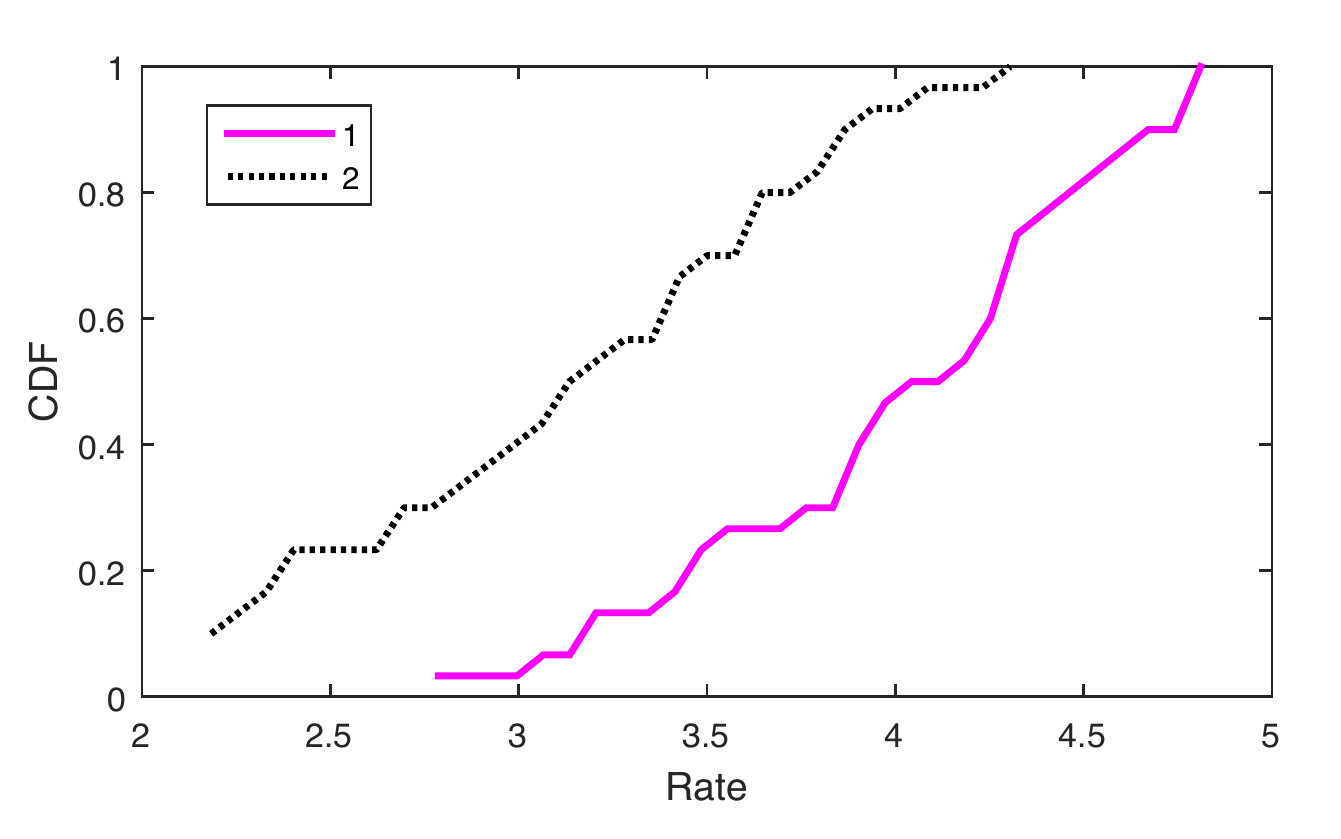}}
\centering
\caption{Comparison with other TPE methods. 
(left) Ergodic sum rate vs. SNR for different TPE methods and degree $J = 3$ [1 = proposed method, 2 = method of \cite{zarei2013low}, 3 = 
method of \cite{kammoun2014linear}]. (right) CDF of the max-min ergodic user rate 
optimized according to the proposed power control method (1) and according to the method in \cite{kammoun2014linear} (2) for $J = 3$ and $\SINR = 20$dB.}
\label{Jplotex2}
\end{figure}

Fig.~\ref{Jplotex2} (left) shows the ergodic sum rate for the various method and the given scattering geometry. 
In addition, since the method of  \cite{kammoun2014linear} explicitly targets the maximization of the minimum ergodic rate, we show in 
Fig.~\ref{Jplotex2} (left) the CDF of the max-min ergodic user rate 
optimized according to the proposed power control method in Section \ref{sec:max-min} 
and according to the method in \cite{kammoun2014linear},  for $J = 3$ and $\SINR = 20$dB. 
Generally speaking, the proposed method outperforms the previous state-of-the art for all tested 
scattering geometries both in terms of sum rate and in terms of max-min rate. The gain of the proposed method over the other 
competing methods depends on the scattering geometry.  Recall that the method of  \cite{kammoun2014linear} handles different 
user channel correlations but, in order to obtain the large-system coefficients in the computation of the TPE weights, 
it requires the iterative solution of a system of coupled matrix fixed-point equations whic is generally too complex for real-time implementation. 
Furthermore, \cite{kammoun2014linear} designs the TPE precoder directly for the DL (i.e., without using UL-DL duality). Hence, a single
set of weights $\wv$ is optimized for all users $k \in [1:K]$. This single set of weights is obtained by maximizing the minimum user SINR. 
In contrast, using UL-DL duality, as in our method, we can optimize a set of TPE coefficients $\wv_k^*$  individually for each user $k$, and 
in addition we can perform some system optimization with respect to the power allocation, as seen in Section \ref{sec:optimization}. 
In contrast, the method of  \cite{zarei2013low} exploits UL-DL duality 
but optimizes only a single set of coefficients $\wv$ according to the minimum total MSE criterion at the output of the UL receiver. 
The large-system limits in \cite{zarei2013low} are obtained using {\em free probability theory} and holds only if 
the users channels have a scaled-identity covariance matrix. 
Hence, in our simulations we have assumed a mismatch of the channel covariance statistics, such that
the method of  \cite{zarei2013low} is applied  {\em assuming} channel covariance $A_k \Id_M$ instead of the true $\Rm_k$. 
This mismatch is unavoidable with the method proposed in \cite{zarei2013low}, and seems to be more severe than the mismatch incurred by our method
replacing the Toeplitz matrices with their circulant approximation, as discussed in Section \ref{sec:large-system-limit}.


\section{Conclusions} \label{sec:conclusions}

In this paper we presented a novel TPE method that outperforms all previously proposed methods in the general non-symmetric case of users with 
arbitrary antenna correlation. In addition, the proposed method is more flexible than other competing methods since, by exploiting UL-DL duality, it can
optimize a set of TPE coefficients for each user. Hence, classical forms of power allocation (e.g., {\em min sum power} and {\em max-min rate}) 
can be easily solved.  Furthermore, the computation of the large-system limiting expressions appearing in the solution of the optimal TPE coefficients 
is obtained by simple direct non-iterative `scalar'' formulas, rather than through a system of coupled matrix fixed-point equations as in the case 
of methods based on {\em deterministic equivalents} \cite{kammoun2014linear,mueller2016linear,lu2016low}. 
In this paper we also provided a novel accurate computation latency analysis of the TPE scheme compared with standard 
``matrix inversion'' RZF, specifically targeted to a highly parallel FPGA hardware architecture. Our analysis shows that
TPE is able to compute the precoding matrix for up to a hundred of LTE resource blocks well within 
a single LTE OFDM symbol period,  as originally assumed in Marzetta's work \cite{Marzetta-TWC10}. 

As a concluding remark, we would like to point out that, in general, the TPE approach 
is able to recover a large fraction of the gap between ConjBF and RZF even for small polynomial degree $J=2,3$. 
However, when users have all the same channel correlation, it may still pay a significant performance gap with respect to
RZF.  However, and perhaps more remarkably, TPE is able to recover virtually the whole gap in a wide range of practically relevant 
SNR if it is used together with some clever scheduling scheme that chooses sets of users arranged in small nearly mutually orthogonal groups.  This scheduling
approach has become popular after the widely referenced work on {\em Joint Space-Division and Multiplexing} (JSDM) 
\cite{adhikary2013joint}. Overall, TPE with JSDM-type scheduling represents a valid alternative to RZF precoding in terms of achievable rate performance.

\appendices

\section{Channel Covariance Model}  \label{sec:geometric-correlation-model}

\begin{figure}[ht]
\centerline{\includegraphics[width=6cm]{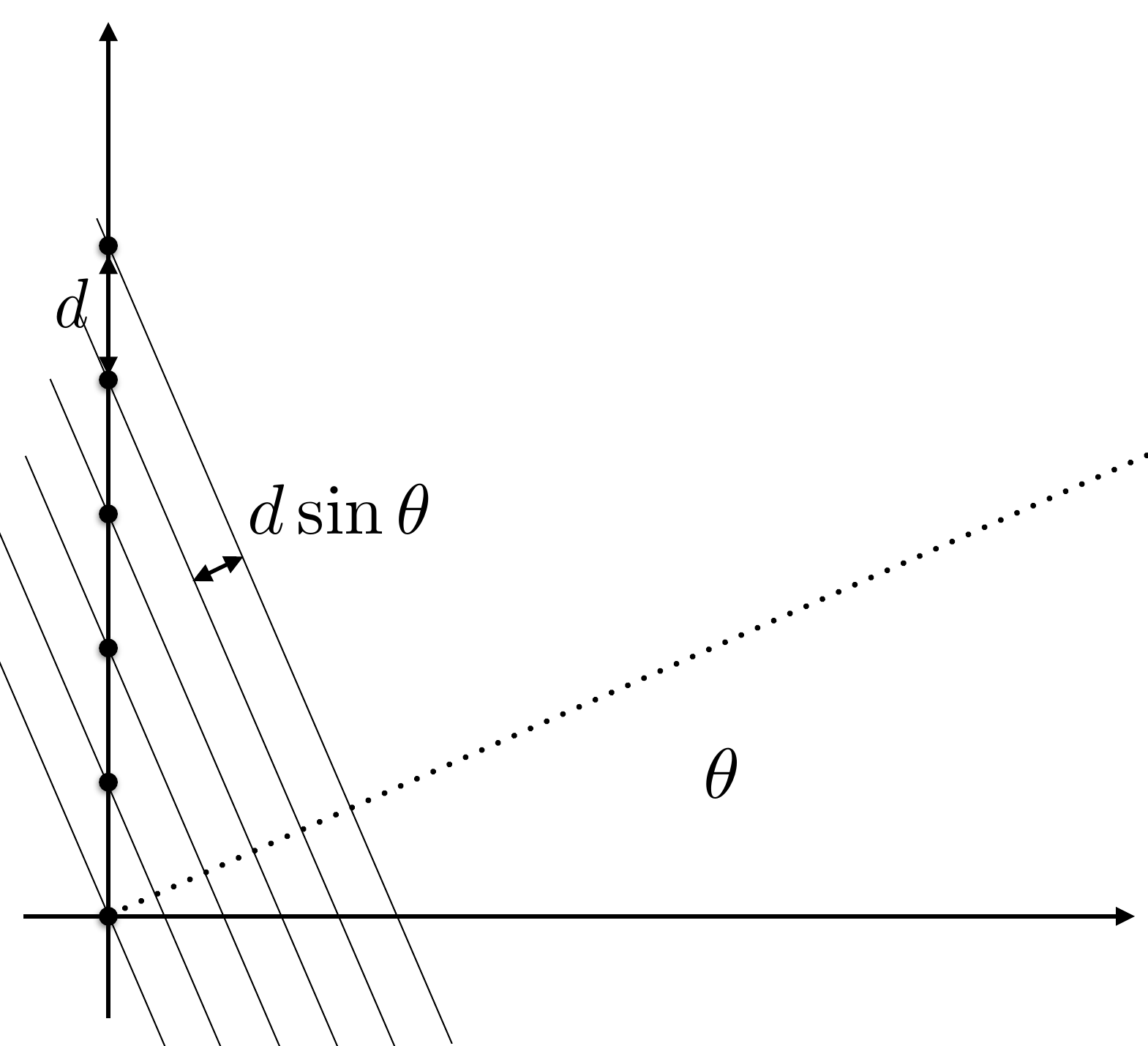}}
\caption{Geometry of a uniformly spaced array with impinging planar wave at an azimuth AoA $\theta$.}
\label{ula}
\end{figure}

In this section we review the channel model that motivates and justify 
the assumptions made at the beginning of Section \ref{sec:large-system-limit}. The user channels in our numerical examples
of Section \ref{sec:results} are also generated according to this model. 
Although the channel model can be found expressed in various equivalent ways in several papers 
(e.g., see \cite{SayeedVirtualBeam2002,Heath2016overview,bajwa2010compressed,adhikary2013joint,gao2015spatially,rao2014distributed}), 
however it is useful to have a concise presentation in a consistent notation with the rest of the paper. 
We assume that the BS is equipped with a Uniformly Space Array (ULA), and that the
scattering field generating the wavefront impinging onto the array are in the array far-field. 
For the geometry of Fig.~\ref{ula}, the array response vector to a planar wavefront coming at AoA
$\theta$ is given by 
\begin{equation} \label{array-response} 
{\bf \asf}(\theta) =  [1, e^{-j2\pi \frac{d}{\lambda} \sin \theta},  \ldots, e^{-j2\pi (M-1) \frac{d}{\lambda} \sin \theta} ]^\transp  
\end{equation}
where $d$ is the spacing between the array elements and $\lambda$ is carrier wavelength. 

The unnormalized channel vectors $\tilde{\hv}_k$ in (\ref{eq:1}) 
can be written as the superposition of array response vectors over the angle of arrival (AoA) $\theta$, weighted by 
scattering coefficients, which are modeled as a Gaussian random field. We have 
\begin{equation} \label{ch-model}
\tilde{\hv}_k = \int {\bf \asf}(\theta) \eta_k(\theta) d\theta,
\end{equation}
where $\eta_k(\theta)$  is an uncorrelated Gaussian random field with autocorrelation function
\begin{equation}  \label{US}
\EE[\eta_k(\theta) \eta_k^*(\theta')] = \rho_k(\theta) \delta(\theta - \theta').
\end{equation}
The {\em angular scattering function} $\rho_k(\theta)$ describes the average energy received from AoA $\theta$. 
Different users are characterized by generally different scattering functions $\rho_k(\theta)$. The channel covariance matrix is given by  
\begin{equation} 
\Rm_k =  \EE[ \tilde{\hv}_k \tilde{\hv}_k^\herm]  
= \int {\bf \asf}(\theta) {\bf \asf}^\herm(\theta) \rho_k(\theta) d\theta. \label{channel-cov}
\end{equation}
Notice that for ULA under uncorrelated scattering and far-field assumption $\Rm_k$ is a Toeplitz matrix with
elements
\begin{equation}
[\Rm_k]_{m,\ell} =  \int e^{-j2\pi \frac{d}{\lambda}(m - \ell) \sin\theta} \rho_k(\theta) d\theta. 
\end{equation}
It is easy to check that the channel strength $A_k$ (as defined in Section \ref{sec:conventional-pc}) is given by 
\begin{equation} 
A_k = \frac{1}{M} \trace(\Rm_k) = \int \rho_k(\theta) d\theta.
\end{equation}
Let $\Fm$ denote the $M \times M$ unitary DFT matrix with $(m,n)$-th element 
$\left [ \Fm \right ]_{n,m} = \frac{1}{\sqrt{M}} e^{-j2\pi \frac{nm}{M}}$ for $m,n \in [0:M-1]$. The array response vector can be expressed in the $\Fm$ basis as
${\bf \asf}(\theta) = \Fm \check{\bf \asf}(\theta)$ where $\check{\bf \asf}(\theta)$ has $m$-th coefficient given by 
\begin{eqnarray} 
\left [ \check{\bf \asf}(\theta) \right ]_m & = & \frac{1}{\sqrt{M}} \sum_{n=0}^{M-1} e^{j2\pi \frac{mn}{M}} e^{-j2\pi \frac{d}{\lambda} \sin(\theta) n} \nonumber \\
& = &  \frac{1}{\sqrt{M}} \frac{\sin(\pi \psi_m(\theta) M)}{\sin(\pi \psi_m(\theta))} e^{-j\pi  \psi_m(\theta) (M-1)}  \label{dirichelet}
\end{eqnarray}
where $\psi_m(\theta) = \frac{d}{\lambda} \sin(\theta) - \frac{m}{M}$. 
We observe that $ \check{\bf \asf}(\theta)$ has peaks for $\psi_m(\theta) \in \ZZ$, and is non-negligible for $|\psi_m(\theta) - z| \leq \frac{1}{M}, \;\;\; z \in \ZZ$. 
It follows that for large $M$ and fixed $\theta$ the coefficient vector $\check{\bf \asf}(\theta)$ tends of have 
large peaks in correspondence of the unique coefficient index $m$ for which $\psi_m(\theta)$ is closets to an integer. We denote such unique index 
as $m(\theta)$, and define the set $\Theta_m = \{ \theta \in [0,2\pi] : m(\theta) = m\}$, where the sets $\Theta_m$ for $m = 0, \ldots, M-1$ are pairwise disjoint. 
Furthermore, from the smoothness of the function $\psi_m(\theta)$ we have that 
$\Theta_m$  is Lebesgue-measurable (in fact, the are finite collections of intervals). 
For what said,  for large $M$ we can approximate
\[ \int \check{\bf \asf}(\theta) \check{\bf \asf}^\herm(\theta) \rho_k(\theta) d\theta  \approx \Lambdam_k \]
where $\Lambdam_k$ is a diagonal matrix with $m$-th diagonal element given by $\Lambda_{m,k} \approx \int_{\Theta_m} \rho_k(\theta)  d\theta$. 
Replacing this approximation into (\ref{channel-cov}) we have that $\Rm_k \approx \Fm \Lambdam_k \Fm^\herm$, 
where the $m$-th eigenvalue yields the received signal energy from the AoA interval $\Theta_m$. The fact that the channel covariance matrix 
can be approximately diagonalized by the unitary DFT matrix justifies replacing $\Rm_k$ with its circulant approximation
in (\ref{circulant-approx}) in order to develop our TPE asymptotic coefficients computation method in Section \ref{sec:large-system-limit}.

\section{Uplink-Downlink duality}   \label{sec:UL-DL-duality}

We review the general UL-DL duality with arbitrary linear vectors as transmitters and receivers, and then particularize to our case.
Consider the normalized UL channel of (\ref{eq:2}), where the input $\xv$ is given by 
$\xv = \sum_{k=1}^K \ev_k x_k$, with $\EE[|x_k|^2] = p_k$ and $\sum_{k=1}^K p_k = K$. 
We can generalize this linearly precoded input in the form 
\[ \xv = \sum_{k=1}^K \uv_k s_k \]
where $\{\uv_k\}$ are arbitrary unit norm vectors and $s_k$ are uncorrelated information symbols.  It follows that this generalized UL channel
is given by 
\begin{equation} 
\yv = \Hm \sum_{k=1}^K \uv_k s_k + \sqrt{\nu} \zv.
\end{equation}
Denote the linear receiver vector for user $k$ by $\vv_k$. Then, the SINR of user $k$ is given by 
\begin{equation}
\SINR_k = \frac{|\vv_k^\herm \Hm \uv_k|^2 p_k/\nu}{1 + \sum_{j\neq k} |\vv_k^\herm \Hm \uv_j|^2 p_j/\nu}.
\end{equation}
We introduce the $K \times K$ matrix $\Phim$ with $(k,j)$-th element given by 
\begin{equation} 
\phi_{k,j} = |\vv_k^\herm \Hm \uv_j|^2. 
\end{equation}
Using the definition of $\Phim$, the SINR can be rewritten as
\begin{equation}  \label{zio}
\SINR_k  = \frac{\phi_{k,k} p'_k}{1 + \sum_{j \neq k} \phi_{k,j} p'_j},
\end{equation}
where we define $p'_k = p_k/\nu$.  It is convenient to introduce the $K \times 1$ vector $\muv$ with $k$-th element
\begin{equation} \label{mu-coef}
\mu_k = \frac{\SINR_k}{(1 + \SINR_k) \phi_{k,k}}, 
\end{equation}
in order to rewrite (\ref{zio}) for all $k$, after some simple algebra, in the matrix-vector form
\begin{equation} \label{zia}
\left ( \Id_K - \diag(\muv) \Phim \right ) \pv' = \muv, 
\end{equation}
where $\pv' = (p'_1, p'_2, \ldots, p'_K)^\transp$.  Notice that  if we fix the UL transmit powers, then (\ref{zio}) 
yields the SINRs of the users for given precoding and receiving vectors $\{\uv_k\}$ and $\vv_k\}$.  

It is well-known (see \cite{Viswanath-Tse-TIT03}) that the collection of SINR values 
\[ \SINR_1 = \Gamma_1, \SINR_2 = \Gamma_2, \ldots, \SINR_K = \Gamma_K \]
is feasible with given $\{\uv_k\}$, $\{\vv_k\}$, and channel matrix $\Hm$, if and only if 
the Perron-Frobenious eigenvalue of the non-negative matrix $\diag(\muv) \Phim$ is strictly less than 1.  
Here ``feasible'' means that there exists a non-negative valued power assignment vector $\pv'$ such that $\SINR_k = \Gamma_k$ 
for all $k \in [1:K]$. In this case, solving (\ref{zia}) for $\pv'$ yields the componentwise minimum power vectors able to achieved 
the given target SINRs.  Namely,the minimum power vector is given by 
\begin{equation}
\pv'_{\min} = \left ( \Id_K - \diag(\muv) \Phim \right )^{-1} \muv, 
\end{equation}  
where, from (\ref{mu-coef}), the coefficients $\mu_k$ are related to the target SINRs by 
$\mu_k = \frac{\Gamma_k}{(1 + \Gamma_k) \phi_{k,k}}$. 

Next, consider the ``dual'' DL, obtained by exchanging the role of transmitters and receivers and ``flipping'' the channel matrix. 
This is given by 
\begin{equation}  \label{dual-dl}
\yv^{\rm dl} = \Hm^\herm \sum_{k=1}^K \vv_k x^{\rm dl}_k + \sqrt{\nu} \zv^{\rm dl}, 
\end{equation}
Notice that $\Hm^\herm$ has dimension $K \times M$, and we are implicitly assuming that $K$ DL users are served 
by linear beamforming, with beamforming vectors $\{\vv_k\}$. User $k$ applies the linear detector $\uv_k$ obtaining the SINR
\begin{eqnarray}
\SINR^{\rm dl}_k & = & \frac{|\uv_k^\herm \Hm^\herm \vv_k|^2 q_k/\nu}{1 + \sum_{j\neq k} |\uv_k^\herm \Hm^\herm \vv_j|^2 q_j\nu} \nonumber \\
& = & \frac{\phi_{k,k} q'_k}{1 + \sum_{j\neq k} \phi_{j,k} q'_j},  \label{ziofa}
\end{eqnarray}
where $q'_k = q_k/\nu$ and $q_k$ is the transmit power for the user $k$ DL data stream. 

Introducing the $K \times 1$ vector $\muv^{\rm dl}$ with $k$-th element
\begin{equation}
\mu^{\rm dl}_k = \frac{\SINR^{\rm dl}_k}{(1 + \SINR^{\rm dl}_k) \phi_{k,k}}, 
\end{equation}
we rewrite (\ref{ziofa}) as done before as
\begin{equation} \label{ziafa}
\left (\Id_K - \diag(\muv^{\rm dl}) \Phim^\transp \right ) \qv' = \muv^{\rm dl}, 
\end{equation}
where $\qv' = (q'_1, q'_2, \ldots, q'_K)^\transp$. 

Now, consider the same assignment of SINRs as in the UL, i.e., for the same values $\Gamma_1, \ldots, \Gamma_K$ we let
\[ \SINR^{\rm dl}_1 = \Gamma_1, \SINR^{\rm dl}_2 = \Gamma_2, \ldots, \SINR^{\rm dl}_K = \Gamma_K. \]
In this case, $\muv^{\rm dl} = \muv$ with components $\mu_k = \frac{\Gamma_k}{(1 + \Gamma_k) \phi_{k,k}}$ and 
the Perron-Frobenious eigenvalues of $\diag(\muv^{\rm dl}) \Phim^\transp$ and of $\diag(\muv) \Phim$ coincide. 
It follows that  if the SINR assignment $\{\Gamma_k\}$ is feasible for the UL it is also feasible for the dual DL. 
Furthermore, solving (\ref{ziafa}) for $\qv'$ yields the componentwise minimum power vectors able to achieved the given target SINRs. 
Namely,the minimum DL power vector is given by 
\begin{equation}  \label{dl-powermin}
\qv'_{\min} = \left ( \Id_K - \diag(\muv) \Phim^\transp \right )^{-1} \muv.
\end{equation}  
Finally, we can easily show that the total sum power of the UL and DL minimum power vectors achieving the same
SINR assignment is equal. In fact
\begin{eqnarray}
\sum_{k=1}^K p'_{\min,k} & = & \onev^\transp \pv'_{\min} \nonumber \\
& = & \onev^\transp \left (\diag(1/\mu_1, \ldots, 1/\mu_K)  - \Phim \right )^{-1} \onev \nonumber \\
& = & \onev^\transp \left (\diag(1/\mu_1, \ldots, 1/\mu_K)  - \Phim^\transp \right )^{-1} \onev \nonumber \\
& = & \onev^\transp \left (\Id  - \diag(\muv) \Phim^\transp \right )^{-1} \muv \nonumber \\
& = & \sum_{k=1}^K q'_{\min,k}.
\end{eqnarray}
As a consequence, in order to optimize the DL of a multiuser MIMO system with linear precoding, 
we can consider a ``virtual'' UL letting $\uv_k = \ev_k$ and  $\vv_k$ given by the unit vectors obtained from the TPE receiver developed 
before, and $p'_k = p_k/\nu = p_k \SNR/\beta = p_k M \SNR/K$, with the sum power constraint
\[ \sum_{k=1}^K p_k' = M \SNR. \]  
The UL SINRs are given by
\begin{equation}  \label{sinr-again}
\SINR_k = \lambda_{k,\max}
\end{equation}
where $\lambda_{k,\max}$ is defined in (\ref{opt-sinr-w}). 
Notice that these SINRs are functions of the channel matrix and of the UL power allocation. Therefore, we can run
run the optimization over the UL powers of any desired objective function of the UL SINRs. 
By duality, it follows that  the resulting (optimized) SINRs $\{\lambda_{k,\max}\}$ can be also achieved in the DL 
by using the same TPE coefficients $\{\wv^*_k\}$  resulting from the UL optimization, 
the corresponding UL receiving vectors $\{\vv_k\}$ as DL precoders,  and the DL power assignment $\qv'_{\min}$ 
obtained from (\ref{dl-powermin}) for the target SINRs $\{\Gamma_k = \lambda_{k,\max}\}$. 

\section{Maximally parallelized QR decomposition algorithm details}  \label{sec:qrh_details}

The following algorithm boxes \ref{alg:HVC}, \ref{alg:HCZ}, \ref{alg:BS} and \ref{alg:QRH} are showing the involved algorithms for the maximally parallelized QR decomposition algorithm using Householder reflections and back substitution for calculation of the classical matrix inverse in the benchmark regularized zero forcing precoder. 

\begin{algorithm}[p]
\caption{Complex Householder Vector Calculation Unit (\texttt{HVC}) \label{alg:HVC}}
\KwIn{\\
$\ma G_k \in \C^{K\times K}$ (matrix of the $k$-th iteration, where $\ma G_1:=\ma G$ is the channel Gramian $\ma G=\ma H^\herm \ma H$)\;
}
\alg{\\
$\sigma = (\ma G_k)^\herm_{k+1:K,k} (\ma G_k)_{k+1:K,k}$ \;
$\alpha= (\ma G_k)^*_{k,k} (\ma G_k)_{k,k}$ \;
$\gamma = (\ma G_k)^\herm_{1:k-1,k} (\ma G_k)_{1:k-1,k}$ \;
$\mu =  1+\sqrt{1+(\gamma+\sigma)/\alpha}$ \;
$\phi_k = \mu (\ma G_k)_{k,k} = \mu\re\{(\ma G_k)_{k,k}\}  + j \mu \im\{(\ma G_k)_{k,k}\}$\;
$\phiv_k = \left(\underbrace{0 \cdots 0}_{k-1 \; \m{times}} \;\phi_k \;(\ma G_k)^T_{k+1:K,k}\right)^T$\;
$\zeta_k = 2/\left(\sigma +  \phi_k  \phi_k^*\right)$\;
}
\KwOut{\\
$\phiv_k \in \C^{K \times 1}$ ($k$-th Householder vector) \;
$\zeta_k \in \R$ ($k$-th Householder normalization scalar) \;
}
\end{algorithm}
\begin{algorithm}[p]
\caption{Complex Householder Column-Zeroing Unit (\texttt{HCZ}) \label{alg:HCZ}}
\KwIn{\\
$\ma G_k \in \C^{K\times K}$ or $\ma Q_k \in \C^{K\times K}$ ($k$-th iteration Gramian or $k$-th iteration reflector aggregator)\;
$\phiv_k \in \C^{K \times 1}$ ($k$-th Householder vector) \;
$\zeta_k \in \R$ ($k$-th Householder normalization scalar) \;
}
\defi{\\
$\ma A_k  \in \C^{K\times K} := \ma G_k$ or $\ma Q_k$\;
$\ma A_{k+1}  \in \C^{K\times K} := \ma G_{k+1}$ or $\ma Q_{k+1}$
}
\alg{\\
$\ve p_k^T = \phiv_k^\herm \ma A_k$ \;
$\ve q_k = \zeta_k \phiv_k$ \;
$\ma B = \ve q_k \ve p_k^T$ \;
$\ma A_{k+1} = \ma A_k - \ma B$ \;
}
\KwOut{\\
$\ma A_{k+1} \in \C^{K\times K}$ (($k+1$)-th iteration Gramian or ($k+1$)-th iteration reflector aggregator)\;
}
\end{algorithm}
\begin{algorithm}[p]
\caption{Complex Back-Substitution Unit (\texttt{BS}) \label{alg:BS}}
\KwIn{\\
$\ma G_K \in \C^{K\times K}$ (last iteration Gramian)\;
$\ma Q_K \in \C^{K\times K}$ (last iteration reflector aggregator) \;
}
\defi{\\
$\ma R \in \C^{K\times K} := \ma G_K$ (upper triangular matrix)\;
$\ma Q \in \C^{K\times K} := \ma Q_K$ (unitary matrix)\;
}
\alg{\\
\For{$k=K$ \KwTo $1$}{
$\map X_{k,:} = \map Q_{k,:} / \map R_{k,k}$ \;
$\map Q_{1:k-1,:} = \map Q_{1:k-1,:} - \map R_{1:k-1,k} \map X_{k,:}$ \;
}
}
\KwOut{\\
$\ma G^{-1} \in \C^{K\times K} := \ma X$ (inverse of the channel Gramian, i.e. $\ma G^{-1} = (\ma H^\herm \ma H)^{-1}$)\;
}
\end{algorithm}
\begin{algorithm}[p]
\caption{Complex QR Decomposition using Householder Reflections Unit (\texttt{QRH})} \label{alg:QRH}
\SetKwFunction{HVC}{HVC}
\SetKwFunction{HCZ}{HCZ}
\SetKwFunction{BS}{BS}
\KwIn{\\
$\ma G\in \C^{K\times K}$ (channel Gramian, i.e. $\ma G=\ma H^\herm \ma H$)\;
}
\defi{\\
$\ma I_K \in \R^{K\times K}:=\diag\{1, 1, \ldots , 1\}$ (Identity matrix)\;
}
\alg{\\
$\ma G_1 := \ma G$ \;
\For{$k=1$ \KwTo $K-1$}{
$\phiv_k , \zeta_k =$ \HVC{$\ma G_k$} \;
$\ma G_{k+1} = $ \HCZ{$\ma G_k, \phiv_k, \zeta_k$} \;
\If{$k=1$}{
$\ma Q_1 = \ma I_K$
}
$\ma Q_{k+1} = $ \HCZ{$\ma Q_k, \phiv_k, \zeta_k$} \;
}
}
$\ma G^{-1} = $ \BS{$\ma G_K, \ma Q_K$} \;
\KwOut{\\
$\ma G^{-1} \in \C^{K\times K}$ (inverse of the channel Gramian, i.e. $\ma G^{-1} = (\ma H^\herm \ma H)^{-1}$)\;
}
\end{algorithm}

For each of the algorithm units the required number of hard resource blocks are given by
\begin{align}
	\chi_{\m{HVC}} & = K \times \m{CCM} + (K-2) \times \m{A} + 1 \times \m{RD} + 1 \times \m{S}\\
	\chi_{\m{HCZ}} & = K^2 \times \m{CM} + 2K \times \m{M} + K^2 \times \m{CA} \\
	\chi_{\m{BKS}} & = 2K \times \m{RD}.
\end{align}
Since the BKS unit can only start its operation when the HVC and HCZ units have completed their task it is 
possible to reuse the resources from these units for the BKS unit. This is the reason why only $2K$ additional division 
blocks are required for the BKS unit. From these equations we can now derive the estimate of sum resource consumption for the QRH unit in terms of DSP blocks
\begin{align}
	\chi_\m{QRH} &= \chi_{\m{HVC}} + \chi_{\m{HCZ}} + \chi_{\m{BKS}} \\
	&= \underbrace{K \times \m{CCM}}_{=K/2 \times \m{CM}} + (K-2) \times \m{A} + 1 \times \m{RD} + 1 \times \m{S}  +  K^2 \times \m{CM} + \underbrace{2K \times \m{M}}_{>K/2 \times \m{CM}} + K^2 \times \m{CA} + 2K \times \m{RD} \\
	&\approx (K^2+K) \times \m{CM} + (K^2+K) \times \m{CA} + (2K+1) \times \m{RD} + 1 \times \m{S} \\ 
	&\approx (K^2+3K) \times \m{CM} + (K^2+K) \times \m{CA} \label{m:nr_step}\\
	&= 4(K^2+3K) \times \m{M} + 2(K^2+3K) \times \m{A} + 2(K^2+K) \times \m{A} \\
	&= 4(K^2+3K) \times \m{M} + 4(K^2+2K) \times \m{A} \\
	&\approx 4(K^2+3K) \times \m{D}. 
\end{align}
Here it was assumed that real division and square root operations are implemented using fast division algorithms like the Newton-Raphson method \cite{koren}. For the sake of simplification it was assumed in \mr{m:nr_step} that the resource requirements for one of these units are similar to a complex multiplier unit. 

\section{Direct calculation of precoded transmit signal vector using TPE and comparison to TPE precoding matrix pre-calculation}\label{sec:dtpep}
We observe that, for the downlink,  an alternative to computing explicitly the beamforming vectors and then computing the precoded signal vector
\[ \xv^{\rm dl} = \sum_{k=1}^K \vv_k s_k \]
consists of directly computing the downlink precoded signal vector $\xv^{\rm dl}$ through a Horner's rule iteration. This has the advantage of avoiding 
products of $K \times K$ matrices, but has the disadvantage that since the information symbols $\{s_k\}$ change at each
OFDM symbol and each subcarrier (i.e., at every channel use in the time-frequency plane of an OFDM system), 
this computation must be repeated for every channel use. Hence, when the same channel coefficients are constant over a large block of signal dimensions
in the time-frequency domain, the direct computation of the precoded signal vector may incur a larger complexity than the explicit computation of the precoding vectors. 
For the sake of completeness, we provide here such computation for a given precoded signal vector. In this case, we can write
\begin{eqnarray}
\xv^{\rm dl} & = & \sum_{k=1}^K \vv_k s_k \\
& = & \sum_{k=1}^K \Hm \sum_{l=0}^J w_{k,l} (\Pm\Gm)^l \ev_k s_k \\
& = & \Hm   \sum_{l=0}^J (\Pm\Gm)^l \sum_{k=1}^K \ev_k w_{k,l}  s_k \\
& = & \Hm   \sum_{l=0}^J (\Pm\Gm)^l \Wm^{(l)} \sv 
\end{eqnarray}
where $\Wm^{(l)}$ was defined before and where $\sv$ is the $K \times 1$ vectors of information symbols to be transmitted in the downlink.
At this point, we define the Horner's type iteration with initialization $\xv^{(J)} = \Wm^{(J)} \sv$ and recursion
\begin{align}
	\xv^{(n)}_k = \Wm^{(n)} \sv   + \Pm\Gm \xv^{(n+1)}, \;\;\; n = J-1, J-2, \ldots, 0 \label{m:dtpe}
\end{align}
such that $\xv^{\rm dl} = \Hm \xv^{(0)}$. A comparison of the computational complexity of this approach with respect to the computation of the 
precoding vectors using (\ref{precoding-vectors-computation}), followed by the computation of the precoded signal vectors using the standard
matrix-vector multiplication $\Vm^\m{tpe}\sv$ for all channel uses, is provided in the following. 

The question to answer is if direct TPE precoding (DTPEP) will provide faster first OFDM symbol output when compared to conventional precoding with pre-computation of the precoding matrix via TPE and subsequent calculation of the transmit vector (TPEP). DTPEP refers to the direct calculation of the transmit signal vector via \mr{m:dtpe}. 

For conventional TPE precoding we will have a latency of 
\begin{align}
	L_\m{TPEP}(U) =  B L_\m{TPE}(U) + sB L_\m{P}(U)
\end{align}
clock cycles in a system with $B$ frequency coherence blocks and $s$ OFDM sub-carriers in one of these blocks. The precoding matrix has to be calculated for $B$ frequency coherence blocks and then the first OFDM symbol which comprises $s B$ sub-carriers has to be precoded. The computational latency for the classical precoder unit (P) will be
\begin{align}
	L_\m{P}(U)  = \begin{cases}\left(N_\m{CM} + \lceil\log_2(K)\rceil N_\m{CA}\right) + \left(\frac{M}{UK} - 1 \right) & \text{if } UK < M\\ 
                                    N_\m{CM} + \lceil\log_2(K)\rceil N_\m{CA} & \text{if } UK \ge M \end{cases}
\end{align}
including pipeling gain. This calculation has to be carried out for every sub-carrier in every OFDM symbol. For example in a LTE system with $\un{20}{MHz}$ channel bandwidth with $B = 100$ resource blocks in the frequency direction and $s=12$ sub-carriers per resource block this calculation has to be carried out $sB=1200$ times for every OFDM symbol, i.e. every $\approx \un{66.7}{\mu s}$. 

For DTPEP we will only  have the pre-computation of the channel Gramian and then direct precoding of the first OFDM symbol comprising $s B$ sub-carriers has to be carried out, yielding a latency of
\begin{align}
	L_\m{DTPEP}(U) =  B L_\m{GC}(U) + sB L_\m{DTPE}(U)
\end{align}
clock cycles. The latency for the direct TPE precoding algorithm (DTPE) will be
\begin{align}
	L_\m{DTPE}(U) = N_\m{CM} + J\left( N_\m{CM} + \lceil \log_2(K) \rceil N_\m{CA} + N_\m{CA} \right) + L_\m{P}(U)
\end{align}
clock cycles. 

\begin{figure}[htb]
	\centering	 \includegraphics[width=12cm]{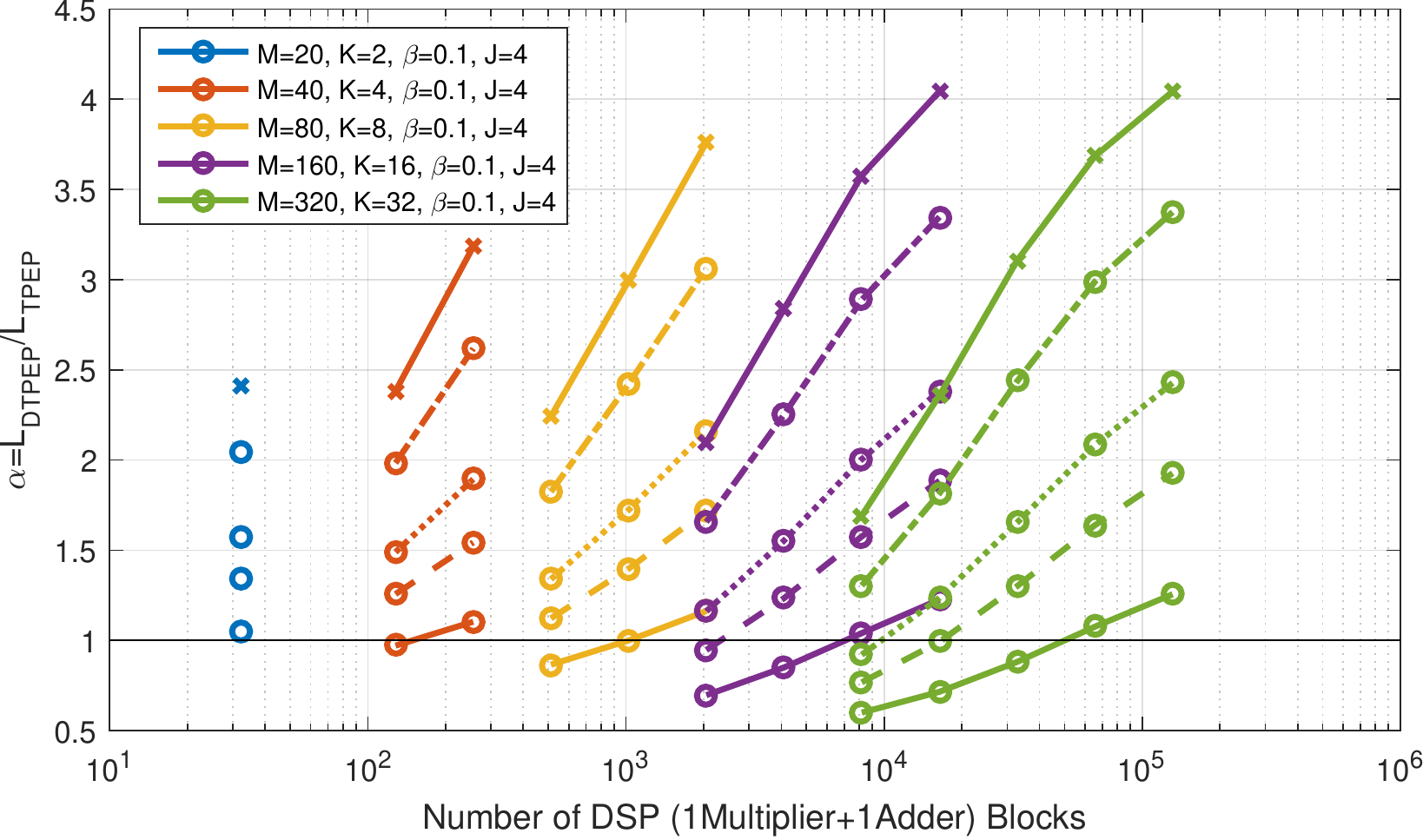}
	\caption{Latency amplification for calculation of first OFDM symbol. Direct TPE precoding and TPE precoding with precoding matrix pre-computation are compared. Different values for the number of OFDM sub-carriers per resource block with $s=2$ (-) $s=4$ (- -) $s=6$ (:) $s=12$ (-.) $s=20$ (x-) are shown.}
	\label{fig:dtpep_vs_tpep}
\end{figure}

Now we can calculate the latency amplification of direct TPE precoding in comparison to classical TPE precoding yielding
\begin{align}\label{m:alpha_dtpep}
	\alpha(U,s) = \frac{L_\m{DTPEP}(U)}{L_\m{TPEP}(U)} = \frac{B L_\m{GC}(U) + sB L_\m{DTPE}(U)}{B L_\m{TPE}(U) + sB L_\m{P}(U)} = \frac{L_\m{GC}(U) + s L_\m{DTPE}(U)}{L_\m{TPE}(U) + s L_\m{P}(U)}.
\end{align}
We can see that the latency amplification is independent of the number of resource blocks but the value of $s$ may play an important role. Figure \ref{fig:dtpep_vs_tpep} shows this latency amplification for multiple values of $s$ (see figure caption) and different values for $M$ and fixed $\beta=0.1$, $J=4$. 

For $M\le 160$ and $s\ge 6$ the latency amplification will be always higher than one, thus direct TPE precoding will be slower than TPE with precoding matrix pre-computation. For an LTE system with $s=12$, $M=160$, $K=16$, $J=4$ and 4096 DSP blocks, TPE using precoding matrix pre-computation will finish with the computation of the first OFDM symbol about 2.2 times earlier than direct TPE precoding. This relationship is simply due to the fact that classical precoding just consists of a single matrix-vector multiplication and direct TPE consists of much more matrix-vector multiplications. And for higher values of $s$ like $s=12$ this just takes longer than just doing one precoding matrix calculation. We can conclude that for LTE's $s=12$ the latency amplification will be always greater than one for any combination of parameter and resource values analyzed here, i.e. TPE using precoding matrix pre-computation will always deliver faster first OFDM symbol output than direct TPE precoding.

\section{TPE coefficients from free probability}  \label{sec:mueller}

In \cite{zarei2013low} an approach to compute the TPE coefficients for DL MU-MIMO precoding is proposed. 
This approach uses also UL-DL duality as in the present work and computes the DL precoder from a ``virtual'' UL channel. 
However, this method is different from ours and it is less general since it critically assumes that the user channels have i.i.d. coefficients. 
Here we review their method expressed with notation and model assumptions compatible with our model in order to enable  a direct comparison.

We let again $M$ denote the number of antennas and $K$ the number of single-antenna users, with $\beta = K/M$. 
The method of \cite{zarei2013low} considers a more restrictive channel matrix model
where the channel covariance matrices are diagonal, i.e., $\Rm_k = A_k \Id$ for all $k = [1:K]$.
The unnormalized channel matrix is therefore given in the form
\begin{equation} 
\tilde{\Hm} = \sqrt{M} \HH  \Am^{1/2} 
\end{equation}
where $\HH$ is i.i.d. of dimension $M \times K$ with elements $\sim \Cc\Nc(0,1/M)$, and $\Am = \diag(A_1, \ldots, A_K)$. 

The dual UL model is given by 
\begin{equation}
\tilde{\yv} = \tilde{\Hm} \tilde{\Pm}^{1/2} \dv + \tilde{\zv}, 
\end{equation}
where $\dv$ is a $K \times 1$ vector of coded symbols with covariance $\EE[ \dv \dv^\herm] = \Id$, 
$\tilde{\zv} \sim \Cc\Nc(\zerov, N_0 \Id)$, and $\tilde{\Pm} = \diag(\tilde{p}_1, \ldots, \tilde{p}_K)$ is the UL power allocation matrix. 
The normalized channel model corresponds to (\ref{eq:2}) and is given by 
\begin{equation}
\yv = \HH \Am^{1/2} \Pm^{1/2} \dv + \sqrt{\frac{\beta}{\SNR}} \zv, 
\end{equation}
where $\zv \sim \Cc\Nc(\zerov, \Id)$, $\SNR = P/N_0$,  and $\Pm = \diag(p_1, \ldots, p_K)$ satisfies the constraint $\sum_k p_k = K$. 
Letting $\Vm = [\vv_1, \ldots, \vv_K]$ the matrix of linear receivers, we have that the UL SINR for user $k$ is given by 
\begin{equation} 
\SINR_k = \frac{|\vv_k^\herm \hv_k|^2 p_k}{\nu \|\vv_k\|^2 + \sum_{j \neq k} |\vv_k^\herm \hv_j|^2 p_j},  
\end{equation}
where as usual we let $\nu = \beta/\SNR$. 

The optimization of the receiver matrix  $\Vm$ is carried out as follows. 
They define the receiver matrix in the form
\begin{equation} \label{ralf-TPE}
\Vm^\herm = \sum_{l=0}^J \omega_l (\HH^\herm \HH)^l \HH^\herm
\end{equation}
and define the estimated user symbols as
\begin{equation}  \label{dhat}
\widehat{\dv} = \Pm^{-1/2} \Am^{-1/2} \Vm^\herm \yv = \Pm^{-1/2} \Am^{-1/2}  \Vm^\herm \HH \Am^{1/2} \Pm^{1/2} \dv +  \sqrt{\nu} \Pm^{-1/2} \Am^{-1/2}  
\Vm^\herm \zv.
\end{equation}
Then, the coefficients $\omegav = (\omega_0, \ldots, \omega_J)^\transp$ are optimized by minimizing the total MSE, defined as
\[ {\sf MSE} = \frac{1}{K} \sum_{k=1}^K \EE [ |d_k - \widehat{d}_k|^2] \]
For convenience, define the combined pathloss and power allocation coefficient matrix $\breve{\Pm} = \Pm\Am$ and rewrite (\ref{dhat}) as
\begin{eqnarray}
\widehat{\dv} & = & \breve{\Pm}^{-1/2} \Vm^\herm \HH \breve{\Pm}^{1/2} \dv +  \sqrt{\nu} \breve{\Pm}^{-1/2} \Vm^\herm \zv.
\end{eqnarray}
Defining the error vector $\ev =  \widehat{\dv} - \dv$ we have
\begin{subequations}
\begin{eqnarray} 
{\sf MSE} & = & \frac{1}{K} \trace ( \EE[ \ev \ev^\herm ] ) \nonumber \\
& = & \frac{1}{K} \trace \left ( \breve{\Pm}^{-1} \Vm^\herm \HH \breve{\Pm} \HH^\herm \Vm \right ) \label{term1} \\
& & - \frac{2}{K} \trace \left ( \Vm^\herm \HH \right )  \label{term2} \\
& & + \frac{\nu}{K} \trace \left ( \breve{\Pm}^{-1} \Vm^\herm \Vm \right ) \label{term3}
\end{eqnarray} 
\end{subequations}
We consider separately the terms (\ref{term1}), (\ref{term2}) and (\ref{term3}), indicated by $T_1$, $T_2$ and $T_3$, respectively. 
For the first term we have
\begin{eqnarray*}
T_1 & = &  \frac{1}{K} \trace \left (\breve{\Pm}^{-1} \Vm^\herm \HH \breve{\Pm} \HH^\herm \Vm \right ) \nonumber \\
& = &  \frac{1}{K} \trace \left ( \left ( \sum_{l=0}^J \omega_l ( \HH^\herm \HH)^{l+1}  \right ) 
\breve{\Pm} \left ( \sum_{l=0}^J \omega_l ( \HH^\herm \HH)^{l+1}  \right )  \breve{\Pm}^{-1}  \right ) 
\end{eqnarray*} 
The large system limit of this term (i.e., for $K \rightarrow \infty$) can be obtained using \cite[Th.1]{zarei2013low}, repeated here for convenience: 

\begin{theorem} \label{mueller}
If $(\Xm, \Ym)$ is asymptotically almost surely free from $(\Cm, \Dm)$, then
\[ \phi \left ( \Xm \Cm \Ym \Dm \right ) = \phi (\Xm) \phi(\Ym) \phi(\Cm \Dm) + \phi(\Xm \Ym) \phi(\Cm) \phi(\Dm) - \phi(\Xm) \phi(\Ym) \phi(\Cm) \phi(\Dm) \]
where $\phi (\cdot)$ is the asymptotic normalized trace operator $\lim_{K \rightarrow \infty} \frac{1}{K} \trace (\cdot)$. 
\hfill $\square$
\end{theorem} 

Here we identify $\Xm = \Ym =  \sum_{l=0}^J \omega_l ( \HH^\herm \HH)^{l+1}$, 
$\Cm = \breve{\Pm}$ and $\Dm = \breve{\Pm}^{-1}$. It follows that, in the limit of large $K$, 
\begin{equation} 
T_1 = (1 - b_1 b_2) \left [ \frac{1}{K} \trace \left ( \sum_{l=0}^J \omega_l (\HH^\herm \HH)^{l+1}  \right ) \right ]^2  + 
b_1 b_2 \frac{1}{K} \trace \left ( \left ( \sum_{l=0}^J \omega_l (\HH^\herm \HH)^{l+1}  \right )^2 \right ) 
\end{equation}
where we define $b_1 = \frac{1}{K} \trace (\breve{\Pm}^{-1})$ and 
$b_2 = \frac{1}{K} \trace (\breve{\Pm})$. 

Let's introduce the following moment: 
\begin{equation}  \label{ralf-moment}
\breve{\rho}_\ell^\infty = \lim_{K \rightarrow \infty} \frac{1}{K} \trace \left ( \HH^\herm \HH \right )^\ell
\end{equation}
Then, using the linearity of the trace operator it is not difficult to show the following:
\begin{equation} 
T_1 = (1 - b_1b_2) \omegav^\transp \breve{\av} \breve{\av}^\transp \omegav + b_1b_2 \omegav^\transp \breve{\Bm} \omegav 
\end{equation}
where we define the $(J+1) \times 1$ vector $\breve{\av}$ with components (for $l \in \{ 0,\ldots, J\}$)
\[ \breve{a}_l = \breve{\rho}_{l+1}^\infty \]
and the $(J+1) \times (J+1)$ matrix $\breve{\Bm}$ with $(l,l')$ elements (for $l,l' \in \{0,\ldots, J\}$)
\[ \breve{B}_{l,l'} = \breve{\rho}^\infty_{l + l' + 2} \]
Next we consider the term $T_2$, for which we can write immediately
\begin{eqnarray}
T_2 & = & \frac{2}{K} \trace \left (\Vm^\herm \HH \right ) \nonumber \\
& =  & \frac{2}{K} \trace \left (  \sum_{l=0}^J \omega_l ( \HH^\herm \HH)^{l+1} \right ) \\
& \rightarrow & 2 \omegav^\transp \breve{\av} 
\end{eqnarray}
Finally, for the term $T_3$ we apply again Th. \ref{mueller} and obtain
\begin{eqnarray}
T_2 & = &  \frac{\nu}{K} \trace \left ( \breve{\Pm}^{-1} \Vm^\herm \Vm \right ) \nonumber \\
& \rightarrow & \nu \frac{1}{K} \trace \left ( \breve{\Pm}^{-1} \right ) \frac{1}{K} \trace \left ( \Vm^\herm \Vm \right ) \nonumber \\
& = & \nu b_1 \sum_{l=0}^J \sum_{l'=0}^J \omega_l \omega_{l'} \frac{1}{K} \trace  \left ( ( \HH^\herm \HH)^{l+ l'+1} \right ) \nonumber \\
& = & \nu b_1 \omegav^\transp \breve{\Cm} \omegav 
\end{eqnarray}
where $\breve{\Cm}$ is the $(J+1) \times (J+1)$ matrix with $(l,l')$ elements (for $l,l' \in \{0,\ldots, J\}$)
\[ \breve{C}_{l,l'} = \breve{\rho}^\infty_{l + l' + 1} \]
It follows that ${\sf MSE} = T_1 - T_2 + T_3$ is the quadratic form in the vector of coefficients $\omegav$
\[ {\sf MSE} = (1 - b_1b_2) \omegav^\transp \breve{\av} \breve{\av}^\transp \omegav + b_1b_2 \omegav^\transp \breve{\Bm} \omegav  - 2 \omegav^\transp \breve{\av} +  \nu b_1 \omegav^\transp \breve{\Cm} \omegav  \]
Differentating and setting the gradient to zero we find the solution
\begin{equation} \label{mueller-coefs}
\omegav^\star = \left [ (1 - b_1b_2) \breve{\av} \breve{\av}^\transp + b_1 b_2 \breve{\Bm} + \nu b_1 \breve{\Cm} \right ]^{-1} \breve{\av}. 
\end{equation}
Finally, we consider the relationship between the asymptotic moment (\ref{ralf-moment}) and 
the theory already developed in Section \ref{sec:large-system-limit}.  From Theorem \ref{theorem-antonia1} in the special case of i.i.d. matrices we have
\begin{eqnarray} 
\breve{\rho}_\ell^\infty  & = & \frac{1}{\beta} \EE[ \xi_{\ell}(\Xsf) ], 
\end{eqnarray}
where $\EE[ \xi_{\ell}(\Xsf)$ is given in (\ref{moments-infinity}) for the case of i.i.d. matrix $\bar{\Hm}$.
Since $\HH$ is i.i.d.,  we can use the explicit expression of the moments of the 
Marchenko-Pastur distribution moments in (\ref{mp-moments}), thus obtaining
\begin{eqnarray} 
\breve{\rho}_\ell^\infty 
& = & \frac{1}{\ell} \sum_{i=0}^{\ell-1} {\ell \choose i} {\ell \choose i + 1} \beta^i \label{ralf-moments-explicit}
\end{eqnarray}
As a sanity check, we notice that expression (\ref{ralf-moments-explicit}) coincides with the one given in \cite[eq. 25]{zarei2013low}. 
This shows that our normalization and mapping of notations is consistent. 

As a final remark, we notice that this method applies only to the case where the channel covariance matrices are diagonal (i.e., 
there is only pathloss $A_k$ but no antenna correlation). However, in order to make comparison with our scheme, we can apply the method
in a  {\em mismatched} way, i.e., the coefficients $\omegav$ are computed as given above, assuming correlation matrix
$A_k \Id$, while in reality the channel vectors are correlated according to $\Rm_k$. 
Notice that the TPE precoder (\ref{ralf-TPE}) uses a polynomial in the channel matrix 
$\HH = \Hm \Am^{-1/2}$, i.e., where the channel vectors have been re-scaled by the inverse of the pathloss amplitude coefficients. 
Hence, when we apply the method of \cite{zarei2013low} to the correlated channel case, we compute the pathloss coefficients as
$A_k = \frac{1}{M} \trace(\Rm_k)$ and we use $\Hm \Am^{-1/2}$ in place of $\HH$ in (\ref{ralf-TPE}). 
This corresponds to  a mismatched precoders that knowns only the channel pathlosses and {\em assumes} uncorrelated 
coefficients.

\section{TPE coefficients from deterministic equivalents}  \label{sec:debbah}

In  \cite{kammoun2014linear} a TPE approach for a multi-cell multiuser downlink system is considered. The approach is quite general since it is able to include explicitly
the effect of channel estimation errors and arbitrary user channel covariance matrices, even without the ``same eigenvectors'' assumption as in our method. 
For the sake of a direct comparison, here we focus on the single-cell case with perfect channel state knowledge, 
and also consider the case of covariances with same eigenvectors, such that the asymptotic TPE coefficients can be computed with the 
large-system limit approach of the present paper and not with the overly complicated method
of deterministic equivalents used in  \cite{kammoun2014linear}. 
Nevertheless, there is a fundamental difference between the method of  \cite{kammoun2014linear} and ours. In \cite{kammoun2014linear} the DL precoder TPE coefficents 
are optimized directly, without exploiting UL-L duality. Hence, a single set of coefficients is optimized for all users. 
Necessarily, the optimization of the users' individual rates becomes a multi-objective optimization problem 
and \cite{kammoun2014linear} takes the approach of optimizing this common set of TPE coefficients 
with respect to the max-min rate criterion.  Hence, while in our case we can choose a desired network utility function with some flexibility, using the approach of
 \cite{kammoun2014linear} one is forced to consider the max-min rate. 

The unnormalized DL channel model is given by 
\begin{equation}  \label{DLmodel}
\yv^{\rm dl} = \tilde{\Hm}^\herm \sum_{k=1}^K \vv_k \tilde{x}^{\rm dl}_k + \zv^{\rm dl},
\end{equation}
where $\zv^{\rm dl}$ is an AWGN vector with components $\sim \Cc\Nc(0,N_0)$ and $\{\vv_k\}$ and unit-energy precoding vectors. 
They consider constant power per downlink stream $\tilde{q}_k = \EE[|\tilde{x}^{\rm dl}|^2] = P/K$. 
They consider a slightly different normalization of the channel matrix, integrating the factor $1/\sqrt{K}$ into the channel matrix. 
However, it is convenient to maintain our normalization, such that we define again $\Hm = \frac{1}{\sqrt{M}} \tilde{\Hm}$, and dividing the received signal 
by $\sqrt{P/\beta}$ we arrive at the normalized DL  model
Recall the DL normalized channel model (\ref{dl}), rewritten here for convenience as
\begin{equation}  \label{DLmodel-normalized}
\yv^{\rm dl} = \Hm^\herm \sum_{k=1}^K \vv_k x^{\rm dl}_k + \sqrt{\nu} \zv^{\rm dl},
\end{equation}
where $\EE[|x^{\rm dl}|^2] = 1$ for all $k = 1\ldots, K$ and, as before, we define 
$\SNR = P/N_0$ and $\nu = \beta/\SNR$. Hence, the SINR for user $k$ takes on the form
\begin{eqnarray}  \label{SINRdl}
\SINR^{\rm dl}_k & = & \frac{|\hv_k^\herm \vv_k|^2}{\nu + \sum_{j\neq k} |\hv_k^\herm \vv_j|^2} \nonumber \\
& = & \frac{|\hv_k^\herm \vv_k|^2}{\nu + \sum_{j=1}^K |\hv_k^\herm \vv_j|^2 - |\hv_k^\herm \vv_k|^2}. 
\end{eqnarray}
The DL linear precoder is given in \cite{kammoun2014linear} in the form
\begin{eqnarray} \label{merouane-TPE}
\Vm & = & \sum_{l=0}^J w_l \left ( \Hm\Hm^\herm \right )^l \Hm \nonumber \\
& = & \sum_{l=0}^J w_l \Gammam^l \Hm, 
\end{eqnarray}
where $\Gammam = \Hm \Hm^\herm$ coincides with $\bar{\Gammam}$ defined in (\ref{new-a-coef}) in the case of uniform powers, i.e., $\Pm = \Id$.

Now, replacing $\Vm = [\vv_1, \ldots, \vv_K]$ in (\ref{merouane-TPE}) into the SINR expression  (\ref{SINRdl}), we obtain
\begin{eqnarray}  \label{SINRdl-TPE}
\SINR^{\rm dl}_k 
& = & \frac{\left | \sum_{l=0}^J w_l\hv_k^\herm \Gammam^l \hv_k \right |^2}{\nu + 
\sum_{j=1}^K \left | \sum_{l=0}^J w_l \hv_k^\herm \Gammam^l \hv_j \right |^2  - 
\left | \sum_{l=0}^J w_l \hv_k^\herm \Gammam^l \hv_k \right |^2} \nonumber \\
& = & \frac{\left | \wv^\transp \widehat{\av}_k \right |^2}{\nu + 
\wv^\transp \widehat{\Bm}_k \wv  -   \left | \wv^\transp \widehat{\av}_k \right |^2}
\end{eqnarray}
where the $(J+1) \times 1$ vector $\widehat{\av}_k = (\widehat{a}_{k,1}, \ldots, \widehat{a}_{k,J})^\transp$
and the $(J+1) \times (J+1)$ matrix $\widehat{\Bm}_k = [\widehat{B}_{k,l,l'} ]$ with elements respectively given by 
\begin{equation} \label{debbah-a}
\widehat{a}_{k,l} = \hv_k^\herm \Gammam^l \hv_k 
\end{equation} 
and 
\begin{equation} \label{debbah-B}
\widehat{B}_{k,l,l'} = \hv_k^\herm \Gammam^{l + l' + 1} \hv_k,
\end{equation}
coincide with the analogous coefficients $\av_k$ in (\ref{new-a-coef}) and $\Bm_k$ in (\ref{new-B-coef}) for the special case of uniform powers.

Furthermore, the sum power constraint is imposed by incorporating the power allocation coefficients directly into the transmit precoding vectors. 
This yields the constraint
\begin{equation} \label{precoding-constraint-Debbah}
\frac{1}{K} \trace \left (\Vm \Vm^\herm \right ) = 1.
\end{equation}
Writing 
\begin{eqnarray}
\Vm \Vm^\herm & = & \sum_{l=0}^J \sum_{l'=0}^J w_l w_{l'} \Gammam^l \Hm \Hm^\herm \Gammam^{l'} \nonumber \\
& = & \sum_{l=0}^J \sum_{l'=0}^J w_l w_{l'} \Gammam^{l+l'+1}
\end{eqnarray}
and taking the trace, we obtain the constraint
\begin{equation} \label{precoding-constraint-Debbah1}
\wv^\transp \widehat{\Cm} \wv = 1
\end{equation} 
where the $(J+1) \times (J+1)$ matrix $\widehat{\Cm}  = [\widehat{C}_{l,l'}]$ has elements
\begin{equation} 
\widehat{C}_{l,l'} = \frac{1}{K} \trace \left (\Gammam^{l+l'+1} \right ).  
\end{equation}
Using (\ref{moments-infinity}), it is immediate to verify  that 
\begin{eqnarray}  \label{antonia-checkC}
\widehat{C}^{\infty}_{l,l'} & = & \lim_{K \rightarrow \infty} \widehat{C}_{l,l'} \nonumber \\
& = & \lim_{M \rightarrow \infty} \frac{1}{\beta} \frac{1}{M} \trace \left (\Gammam^{l+l'+1} \right ) \nonumber \\
& = & \frac{1}{\beta} \EE[ \xi_{l+l'+1}(\Xsf) ] 
\end{eqnarray}
where $\EE[ \xi_{\ell}(\Xsf) ]$ is exactly given in (\ref{moments-infinity}). This is due to the fact that since $\Gammam$ and $\Gammam_k = \Hm_k \Hm_k^\herm$ differ just by the rank-1 term $\hv_k \hv_k^\herm$, their asymptotic normalized trace, as well as that of all finite powers, coincide.

Also, we observe that the coefficients in (\ref{debbah-a}) and in (\ref{debbah-B}) are instances of the quadratic form 
$\rho_{k,\ell} = \hv_k^\herm \bar{\Gammam}^\ell \hv_k$ 
for the special case of uniform powers $\Pm = \Id$,  the calculation of which in the asymptotic large-system regime and general power allocation 
has been already treated in Section \ref{sec:large-system-limit}.

Once the asymptotic coefficients $\widehat{\av}_k, \widehat{\Bm}_k$ and $\widehat{\Cm}$ are obtained as described above,  
since a single set of TPE coefficients $\wv$ must be ``good'' for all users $k = 1, \ldots, K$, the authors of \cite{kammoun2014linear} 
propose to maximize the minimum user rate. 

The corresponding optimization problem in Epigraph form is
\begin{eqnarray}
\mbox{maximize}_{\wv, \xi} & & \xi  \label{maxmin1} \\
\mbox{subject to} & & \wv^\transp \widehat{\Cm} \wv = 1, \nonumber \\
\forall k = 1, \ldots, K & & 
 \frac{\left | \wv^\transp \widehat{\av}_k \right |^2}{\nu + 
 \wv^\transp \widehat{\Bm}_k \wv  -   \left | \wv^\transp \widehat{\av}_k \right |^2}  \geq \xi  \nonumber
\end{eqnarray}
This problem is known to be NP-hard. Then, \cite{kammoun2014linear}  suggests the usual semi-definite relaxation that consists of first rewriting (\ref{maxmin1}) as
\begin{eqnarray}
\mbox{maximize}_{\wv, \xi} & & \xi  \label{maxmin2} \\
\mbox{subject to} & & \trace ( \wv \wv^\transp \widehat{\Cm} ) = 1, \nonumber \\
\forall k = 1, \ldots, K & & 
 \frac{\widehat{\av}_k^\transp \wv \wv^\transp \widehat{\av}_k }{\nu + 
\trace ( \wv  \wv^\transp \widehat{\Bm}_k) -   \widehat{\av}_k^\transp \wv \wv^\transp \widehat{\av}_k }  \geq \xi  \nonumber
\end{eqnarray}
and then replace the rank-1 symmetric positive semidefinite matrix $\wv \wv^\transp$ with the symmetric 
positive demidefinite matrix $\Wm \succeq 0$. The resulting semi-definite relaxation yields the problem
\begin{eqnarray}
\mbox{maximize}_{\Wm, \xi} & & \xi  \label{maxmin3} \\
\mbox{subject to} & & \trace ( \Wm \widehat{\Cm} ) = 1, \nonumber \\
& & \Wm \succeq 0 \nonumber \\
\forall k = 1, \ldots, K & & 
 \frac{\widehat{\av}_k^\transp \Wm \widehat{\av}_k }{\nu + 
\trace ( \Wm \widehat{\Bm}_k) -   \widehat{\av}_k^\transp \Wm \widehat{\av}_k }  \geq \xi  \nonumber
\end{eqnarray}
Finally, they notice that (\ref{maxmin3}) can be solved by a line search (e.g., using the bisection method) on the variable $\xi \in [0, \xi_{\max}]$, by considering 
for each fixed value $\xi$ the power minimization problem
\begin{eqnarray}
\mbox{minimize} & & \trace ( \Wm \widehat{\Cm} )   \label{feasibility} \\
\mbox{subject to} & & \Wm \succeq 0 \nonumber \\
\forall k = 1, \ldots, K & & 
\frac{1 + \xi}{\xi} \widehat{\av}_k^\transp \Wm \widehat{\av}_k  - \trace ( \Wm \widehat{\Bm}_k) \geq 
\nu  \nonumber
\end{eqnarray}
Let $\Wm_\xi$ denote the solution of (\ref{feasibility}). Let $\lambda_{\max}(\Wm_\xi)$ denote the maximum eigenvalue of $\Wm_\xi$ and
$\wv_\xi$ denote the corresponding (unit norm) eigenvector. 
Notice that if $\Wm_\xi$ has rank 1, then $\Wm_\xi = \lambda_{\max}(\Wm_\xi) \wv_\xi \wv^\herm_\xi$. Otherwise,  the candidate precoding coefficient vector
is given by $\sqrt{\lambda_{\max}(\Wm_\xi)}  \wv_\xi$, yielding the best rank-1 approximation of $\Wm_\xi$. 
If $\lambda_{\max}(\Wm_\xi) \wv^\transp_\xi \widehat{\Cm} \wv_\xi \leq 1$ then the upper interval in the bisection is selected. If 
$\lambda_{\max}(\Wm_\xi) \wv^\transp_\xi \widehat{\Cm} \wv_\xi > 1$ or the problem is infeasible, 
then the lower interval is selected. Notice also that (\ref{feasibility}) is a semidefinite program, that can be 
solved with standard semidefinite programming tools. 
As already discussed before, the bisection search can be initialized with the interval 
$[0, \xi_{\max}]$ with $\xi_{\max} = \SNR \max_k \trace(\Rm_k)$.

{\footnotesize
\bibliographystyle{IEEEtran}
\bibliography{TPE-refs-AB}
}

\end{document}